\documentclass[pdflatex,sn-mathphys-num]{sn-jnl}

\usepackage{graphicx}%
\usepackage{multirow}%
\usepackage{amsmath,amssymb,amsfonts}%
\usepackage{amsthm}%
\usepackage{mathrsfs}%
\usepackage[title]{appendix}%
\usepackage{xcolor}%
\usepackage{textcomp}%
\usepackage{manyfoot}%
\usepackage{booktabs}%
\usepackage{chemformula}
\usepackage{algorithm}%
\usepackage{algorithmicx}%
\usepackage{algpseudocode}%
\usepackage{listings}%
\usepackage{geometry}
\theoremstyle{thmstyleone}%
\theoremstyle{thmstyletwo}%

\theoremstyle{thmstylethree}%

\begin{document}

\title[Article Title]{From inconsistency to decision: explainable operation and maintenance of battery energy storage systems}


\author[1,4,5]{\fnm{Jingbo} \sur{Qu}}
\equalcont{These authors contributed equally to this work.}

\author*[2,3]{\fnm{Yijie} \sur{Wang}}
\email{yijiewang@sjtu.edu.cn}
\equalcont{These authors contributed equally to this work.}

\author[2]{\fnm{Yujie} \sur{Fu}}

\author[2]{\fnm{Putai} \sur{Zhang}}

\author*[4,5]{\fnm{Weihan} \sur{Li}}
\email{Weihan.Li@isea.rwth-aachen.de}

\author*[2,3]{\fnm{Mian} \sur{Li}}
\email{mianli@sjtu.edu.cn}

\affil*[1]{\orgdiv{Global College},
\orgname{Shanghai Jiao Tong University},
\orgaddress{\street{800 Dongchuan Road, Minhang District},
\city{Shanghai}, \postcode{200240}, \country{China}}}

\affil[2]{\orgdiv{Global Institute of Future Technology},
\orgname{Shanghai Jiao Tong University},
\orgaddress{\street{800 Dongchuan Road, Minhang District},
\city{Shanghai}, \postcode{200240}, \country{China}}}

\affil[3]{\orgdiv{Key Laboratory of Urban Complex Risk Control and Resilience Governance},
\orgname{Shanghai Jiao Tong University},
\orgaddress{\street{800 Dongchuan Road, Minhang District},
\city{Shanghai}, \postcode{200030}, \country{China}}}

\affil[4]{\orgdiv{Center for Ageing, Reliability and Lifetime Prediction of Electrochemical and Power Electronic Systems (CARL)},
\orgname{RWTH Aachen University},
\orgaddress{\street{Campus-Boulevard 89},
\city{Aachen}, \postcode{52074}, \country{Germany}}}

\affil[5]{\orgdiv{Institute for Power Electronics and Electrical Drives (ISEA)},
\orgname{RWTH Aachen University},
\orgaddress{\street{Campus-Boulevard 89},
\city{Aachen}, \postcode{52074}, \country{Germany}}}


\abstract{Battery Energy Storage Systems (BESSs) are increasingly critical to power-system stability, yet their operation and maintenance (O\&M) remain dominated by reactive, expert-dependent diagnostics. While cell-level inconsistencies provide early warning signals of degradation and safety risks, the lack of scalable and interpretable decision-support frameworks prevents these signals from being effectively translated into operational actions. Here we introduce an inconsistency-driven O\&M paradigm for large-scale BESSs that systematically transforms routine monitoring data into explainable, decision-oriented guidance. The proposed framework integrates multi-dimensional inconsistency evaluation with large language model-based semantic reasoning to bridge the gap between quantitative diagnostics and practical maintenance decisions. Using eight months of field data from an in-service battery system comprising 3,564 cells, we demonstrate how electrical, thermal, and aging-related inconsistencies can be distilled into structured operational records and converted into actionable maintenance insights through a multi-agent framework. The proposed approach enables accurate and explainable responses to real-world O\&M queries, reducing response time and operational cost by over 80\% compared with conventional expert-driven practices. These results establish a scalable pathway for intelligent O\&M of battery energy storage systems, with direct implications for reliability, safety, and cost-effective integration of energy storage into modern power systems.}

\keywords{Battery energy storage system,  Operation and maintenance, Multi-agent system, Large language model, Inconsistency evaluation }



\maketitle

\begin{figure}[h]
\centering
\includegraphics[page=1, width=\linewidth]{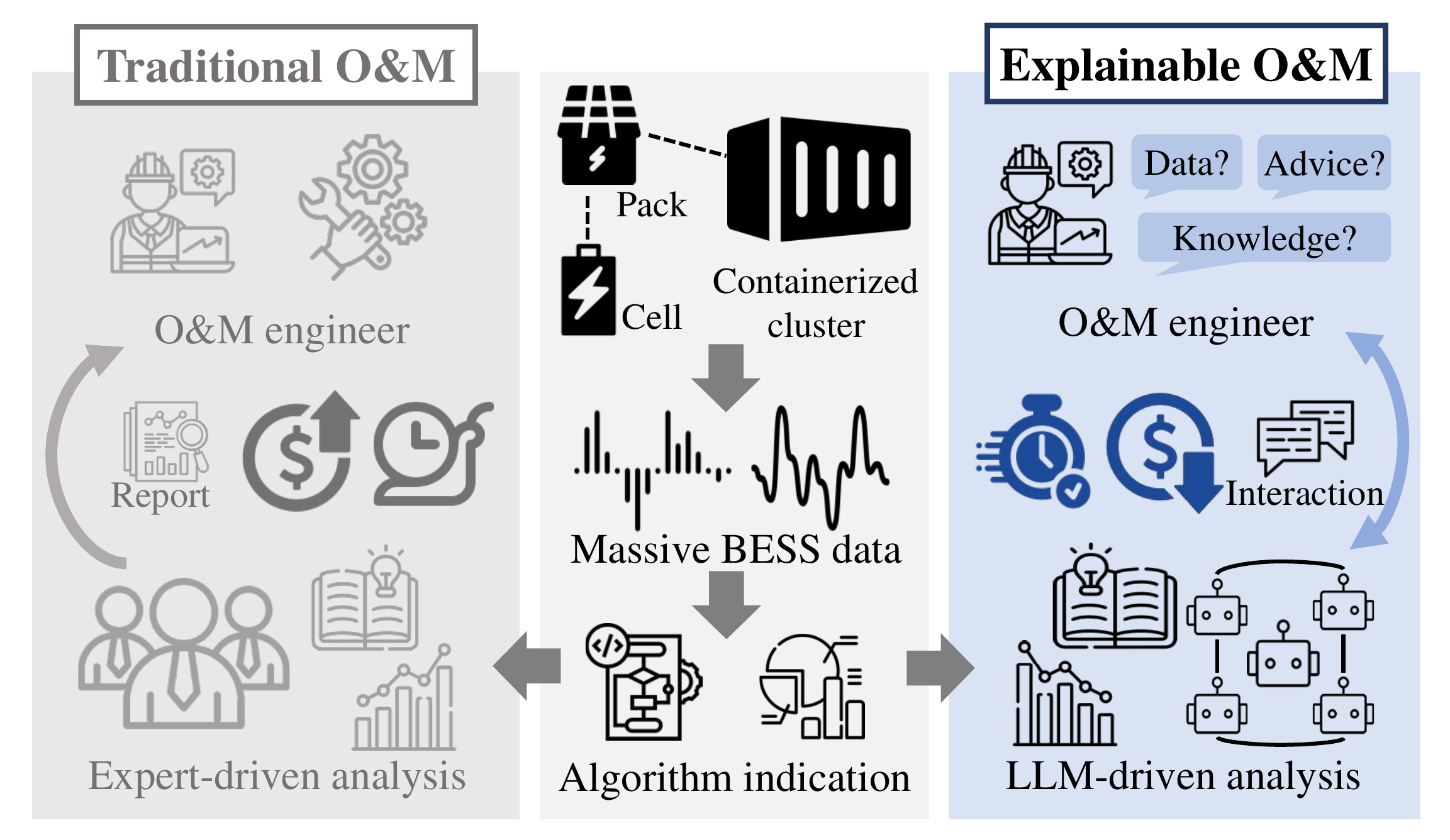}
\caption{\textbf{Motivation of the proposed framework in BESS applications.} In BESS applications, cells are assembled into packs, and packs are further integrated into clusters, generating large volumes of measurement data. Conventional O\&M typically depends on domain experts to interpret algorithm outputs and produce reports that guide O\&M engineers. However, this workflow is largely one-directional and often involves high time and labor costs.
In contrast, the proposed O\&M framework replaces expert-driven reporting with an LLM-based analysis that integrates domain knowledge with algorithm indicators to deliver expert-level insights. This enables interactive, bidirectional communication and real-time responses, allowing engineers to quickly obtain tailored guidance for various BESS O\&M tasks. As a result, the framework improves O\&M efficiency while reducing overall cost.}
\label{fig1}
\end{figure}

\section{Introduction}\label{sec1}
Driven by the global transition to a green economy and the rapid advancement of new energy technologies, Energy Storage Systems (ESSs) have emerged as critical pillars of modern energy infrastructure \cite{intro1}. Lithium-ion Battery Energy Storage Systems (BESSs) have become the dominant choice for large-scale deployments, owing to their high energy density, high conversion efficiency, and low self-discharge rates \cite{intro4}. With the integration of modern active distribution networks, BESSs are organized into hierarchical energy systems, serving as flexible resources that support grid-level optimization and secure operation \cite{intro5}.

Despite these advantages, BESSs face persistent safety and reliability challenges. To achieve the required capacity, typically hundreds of gigawatt-hours, millions of cells are connected in series and parallel. However, inherent inconsistencies across cells, manifesting as variations in voltage, state of charge (SOC), temperature, and state of health (SOH), pose potential risks to the long-term safety of BESS. \cite{knowledge_dataset7, knowledge_dataset10, intro6}. Unlike catastrophic faults, inconsistency represents a "sub-health" state during normal operation, acting as an early warning sign of potential failures \cite{intro7}. If neglected, these minor deviations can propagate to the pack level, causing imbalanced power output, accelerated degradation, and even thermal runaway \cite{intro8}. Importantly, battery inconsistency is not an episodic anomaly triggered by extreme conditions, but a long-term and cumulative phenomenon embedded in daily operation \cite{intro13}. It arises from the coupled effects of manufacturing variability, operational stress, and aging processes, making it particularly suitable for continuous monitoring and proactive management within the operation and maintenance (O\&M) framework of BESSs. From an operational perspective, the primary challenge is therefore not detecting inconsistency itself, but deciding how observed inconsistencies should trigger maintenance or operational interventions.

Numerous approaches have been proposed for battery inconsistency evaluation, which can be broadly categorized into model-based, learning-based, and statistical methods. Representative studies exploit equivalent circuit models \cite{intro9, intro10}, neural network-based reconstruction errors \cite{intro11, intro_encoder}, or handcrafted statistical features \cite{intro12, thermal_incon1} to quantify abnormal behaviors. Despite their effectiveness in detecting abnormal behaviors, existing inconsistency evaluation methods largely terminate at numerical indicators, leaving the translation from metrics to operational decisions dependent on manual expert interpretation. In practical BESS O\&M, however, such metrics represent only an intermediate step rather than the final objective. Industrial O\&M pipelines require these results to be interpreted, contextualized, and ultimately transformed into actionable decisions, such as maintenance prioritization or operational adjustments. Current approaches fail to close this loop between data-driven inconsistency evaluation and scalable, intelligent O\&M. The translation from inconsistency metrics to maintenance actions is still predominantly performed manually by domain experts, relying on experience-driven reasoning that is difficult to standardize, automate, or scale. As BESS deployments continue to expand, this metric-to-decision disconnection becomes a critical bottleneck, limiting both the efficiency and reliability of intelligent O\&M. To date, a systematic and scalable framework that transforms battery inconsistency evaluation into explainable, decision-oriented O\&M for real-world BESSs has not been established.

Recent advances in Large Language Models (LLMs) offer a promising pathway to address this fundamental limitation. LLMs exhibit strong capabilities in semantic understanding, knowledge representation, and logical reasoning, enabling them to interpret heterogeneous analytical outputs within an operational and domain-specific context \cite{intro13}. Such capabilities are particularly relevant for BESS O\&M, where heterogeneous diagnostic outputs must be interpreted within an operational and domain-specific context to support timely decision-making. By integrating LLMs into BESS O\&M pipelines, potentially through structured multi-agent architectures \cite{intro14} and Retrieval Augmented Generation (RAG) \cite{intro15, intro16, intro17}, it becomes possible to systematically encode expert knowledge, explain the implications of inconsistency evaluation results, and generate interpretable, actionable maintenance recommendations \cite{LLMBattery, meth_1}. While LLM-based approaches have been explored in related battery research tasks \cite{llmsoh, llmsoc, llmmaterial}, their application to bridging the gap between inconsistency evaluation and practical BESS O\&M decision-making remains largely unexplored. This work aims to leverage LLM-driven reasoning to close this loop, moving beyond metric-level analysis toward scalable, intelligent, and decision-oriented BESS O\&M.

Motivated by the above observations, this paper proposes an end-to-end explainable O\&M framework for BESS applications that leverages LLM-based semantic reasoning to transform inconsistency evaluation results into actionable maintenance decisions, as shown in Fig. \ref{fig1}. Rather than proposing new inconsistency metrics, this work focuses on closing the loop between inconsistency evaluation and practical O\&M decision-making. The framework is validated using field operating data from 3,564 cells within a containerized battery cluster deployed in an in-service BESS. It contains eight months of operating data encompassing more than 400 distinct operations. The data is publicly released in this work to support reproducibility and future research in BESS operation and maintenance.

The proposed framework can be summarized into three key components: inconsistency evaluation, knowledge formation, and LLM-driven decision-making. First, inconsistency evaluation is performed using dedicated analytical algorithms, and the resulting outputs are organized into a structured record dataset. In parallel, a domain knowledge dataset is established to encapsulate expert knowledge and operational rules. Based on these two datasets, a multi-agent system is constructed to conduct semantic reasoning, interpret inconsistency evaluation outcomes, and provide online, explainable responses to user queries with expert-level maintenance recommendations. By tightly integrating algorithmic modeling, knowledge curation, and LLM-based inference within a unified pipeline, the proposed framework effectively closes the loop between data-driven inconsistency evaluation and practical O\&M execution. Validation results demonstrate that the proposed approach can accurately interpret system behaviors, reveal underlying inconsistency mechanisms, and significantly reduce reliance on manual expert analysis. Compared with traditional expert-driven O\&M modes, the average labor cost per query is reduced by 98.7\%, while the response time is shortened by 81.4\%. Overall, the proposed framework establishes a scalable and intelligent paradigm for BESS O\&M, offering substantial potential for improving operational efficiency and economic performance in large-scale energy storage deployments.

\section{Results}\label{sec2}
\subsection{Framework overview}
\begin{figure}[h]
\centering
\includegraphics[page=3, width=\linewidth]{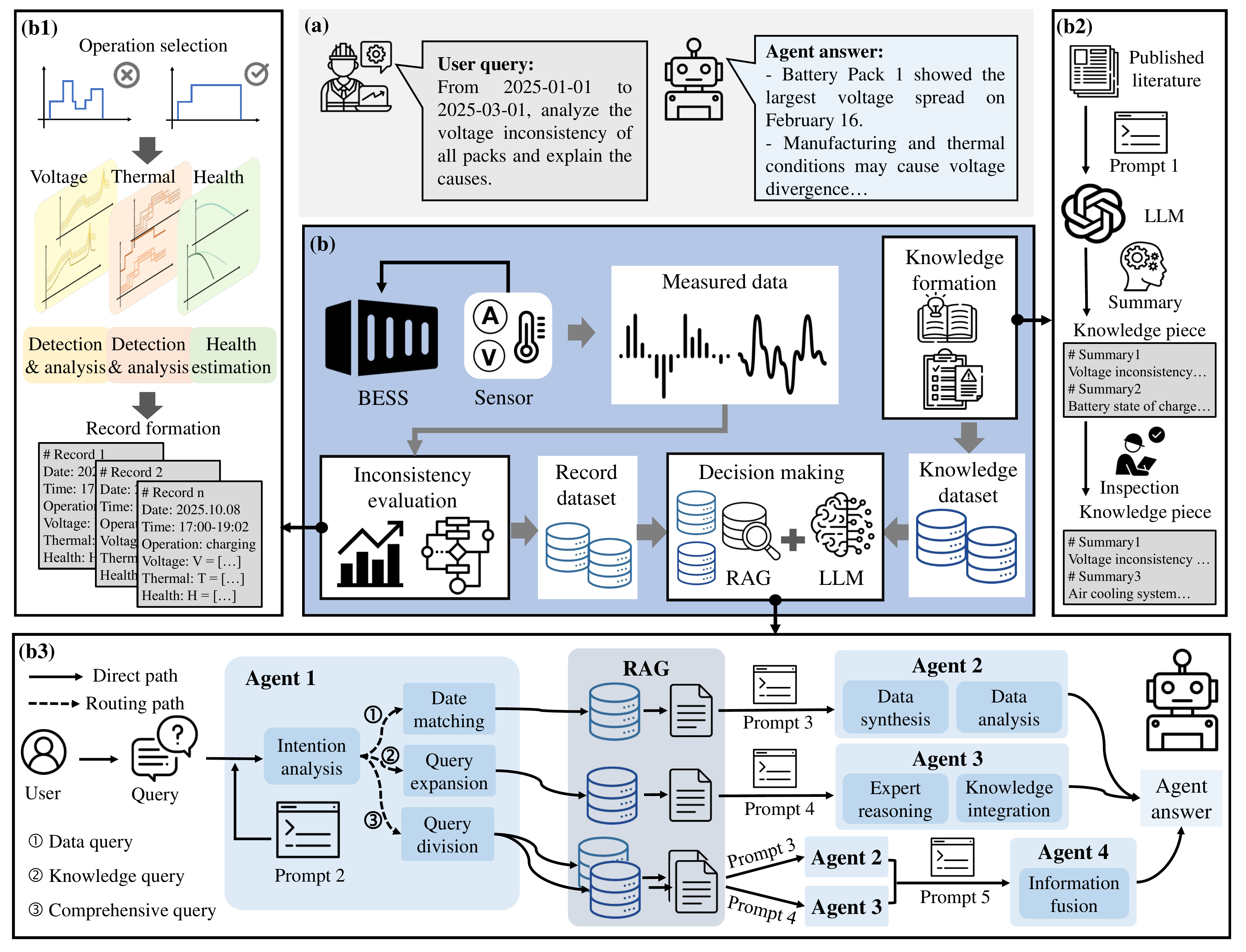}
\caption{\textbf{Architecture of the proposed explainable O\&M framework for BESS applications.} 
\textbf{a} Illustration of a routine query-answering scenario, where an O\&M engineer interacts with the proposed framework to obtain data-driven and explainable maintenance insights. 
\textbf{b} Core paradigm of the proposed O\&M framework (highlighted in deep blue). Measurement data collected from BESS sensors are processed through inconsistency evaluation algorithms to construct a record dataset, while domain-specific expert knowledge is curated to form a knowledge dataset. These two datasets jointly support decision-making by integrating retrieval-augmented generation (RAG) with large language model (LLM)-based inference. 
\textbf{b1} Inconsistency evaluation pipeline. Standard operating conditions are first selected based on predefined criteria, followed by multi-perspective inconsistency assessment from electrical (voltage), thermal, and health-related viewpoints. The resulting indicators are recorded to form the record dataset. 
\textbf{b2} Knowledge formation pipeline. Domain knowledge is distilled from published literature using an LLM guided by carefully designed prompts. The generated knowledge pieces are subsequently reviewed by human experts before being incorporated into the knowledge dataset. 
\textbf{b3} LLM- and RAG-driven decision-making process. User queries are first analyzed by an intent-analysis agent to determine the appropriate routing strategy. For data-oriented queries, relevant records are retrieved from the record dataset and further synthesized through analytical processing. For knowledge-oriented queries, relevant content is retrieved from the knowledge dataset and processed via expert reasoning. For comprehensive queries, both workflows are executed in parallel, followed by information fusion to generate the final response.}
\label{fig2}
\end{figure}

This study proposes a multi-agent–driven explainable O\&M framework for BESS applications, with a particular focus on battery-pack inconsistency analysis. As illustrated in Fig. \ref{fig2}a, the proposed framework enables direct query-answering interactions between O\&M engineers and the intelligent system, allowing users to rapidly obtain system-level insights and context-aware maintenance guidance during real-world operation.

The overall architecture of the framework is shown in Fig. \ref{fig2}b. Operational measurement data are continuously collected from in-service BESS stations and processed through inconsistency evaluation algorithms, which compress high-dimensional time-series data into a structured record dataset. In parallel, domain-specific knowledge related to battery inconsistency mechanisms, operational constraints, and maintenance practices is curated to construct a knowledge dataset. During online operation, user queries are addressed by jointly retrieving and synthesizing information from these two datasets, enabling real-time decision support while maintaining offline updates of both data and knowledge bases. 

The proposed framework consists of three tightly integrated components: construction of an inconsistency record dataset, formation of a domain knowledge dataset, and large language model-based decision-making. As illustrated in Fig. \ref{fig2}(b1), the record dataset is constructed by first selecting representative standard operating conditions, followed by multi-perspective inconsistency evaluation from electrical, thermal, and aging-related viewpoints. The resulting indicators provide a compact yet interpretable representation of the operational state of battery packs. In parallel, the knowledge dataset is established by distilling expert knowledge from relevant literature using an LLM-driven strategy, as shown in Fig. \ref{fig2}(b2). Each knowledge unit is reviewed by domain experts and embedded using semantic keys to support accurate retrieval.

Together, the record and knowledge datasets support an intelligent, multi-agent decision-making workflow, illustrated in Fig. \ref{fig2}(b3). Upon receiving a user query, the system first performs intent analysis to determine whether the request is data-oriented, knowledge-oriented, or comprehensive. Depending on the query type, the framework retrieves relevant operational records, domain knowledge, or both, and synthesizes the information into an explainable response. By explicitly linking observed inconsistency patterns with underlying mechanisms and maintenance implications, the framework enables engineers to move beyond metric-level diagnostics toward decision-oriented O\&M actions.

Rather than introducing new inconsistency metrics, the proposed framework focuses on closing the long-standing gap between inconsistency evaluation and practical O\&M execution. By integrating structured operational data with semantic reasoning in a unified workflow, it establishes a scalable and interpretable paradigm for intelligent O\&M in large-scale battery energy storage systems.

\subsection{Dataset analysis}

\begin{figure}[h]
\centering
\includegraphics[page=2, width=\linewidth]{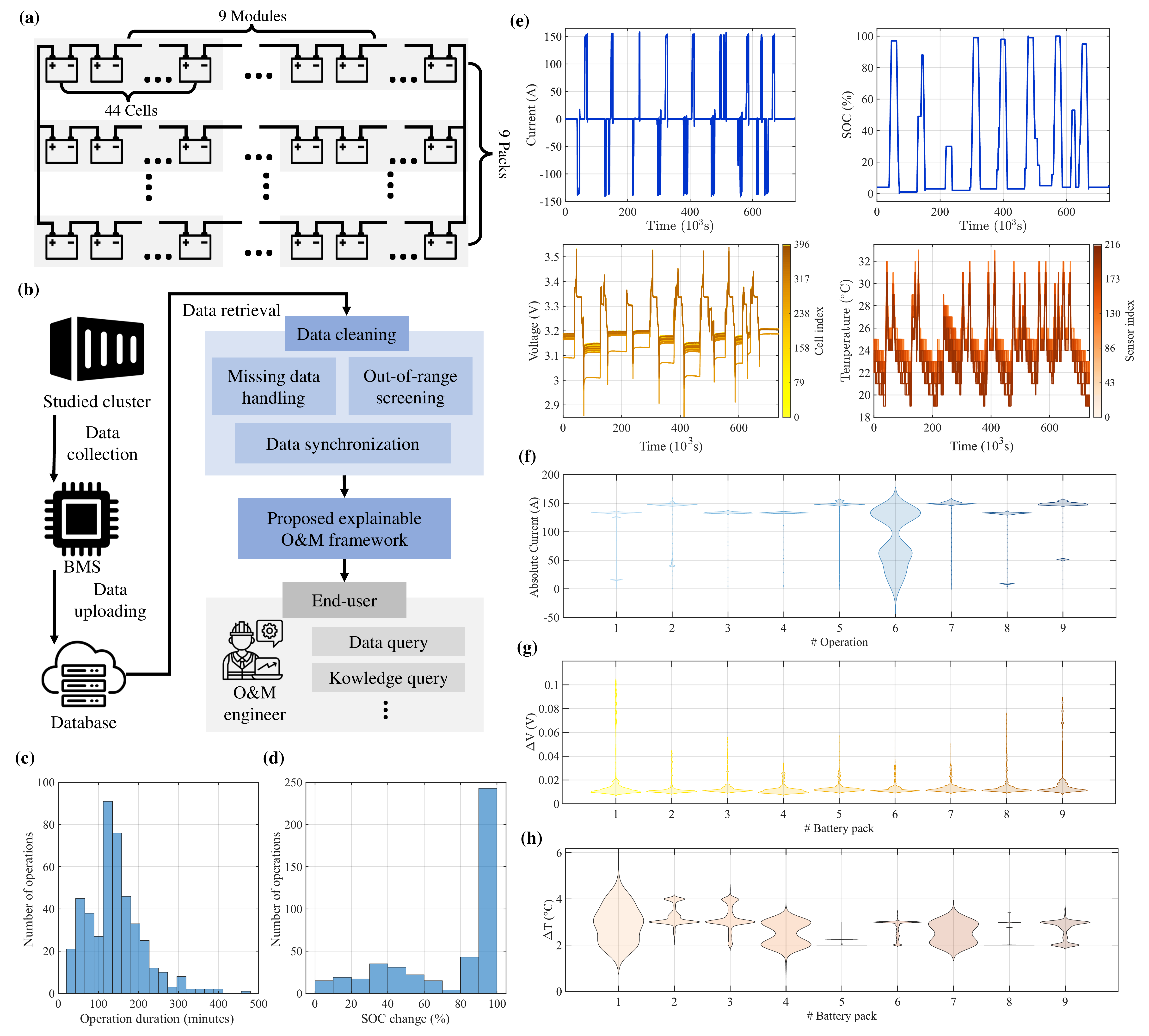}
\caption{\textbf{Data acquisition and dataset analysis.} \textbf{a} Structure of the studied BESS cluster. It consists of nine packs connected in parallel. Each pack contains nine battery modules connected in series, and each module comprises 44 series-connected cells. \textbf{b} BESS data flow. Measurement data from the studied cluster are collected and transmitted to the local Battery Management System (BMS), and then uploaded to a cloud database for storage. When BESS analysis is required, the data are retrieved from the cloud database and preprocessed through data cleaning, including missing-data handling, data screening, and data synchronization. The cleaned data are subsequently fed into the proposed O\&M framework to provide interactive decision support for O\&M engineers and to address various maintenance-related queries. \textbf{c} Statistics on operation duration across the dataset.
\textbf{d} Statistics on SOC change from battery pack 1 across operations in the dataset. \textbf{e} Sample sequences of the measured data from the studied BESS, including current, SOC, voltage, and temperature measurements. \textbf{f} Distribution of current magnitude from operation 1 to 9 in October 2024. \textbf{g} Distribution of the cell-level voltage spread within all packs during Operation 1 in October 2024. \textbf{h} Distribution of the sensor temperature spread within all packs during Operation 1 in October 2024.}
\label{fig3}
\end{figure}

The proposed framework is validated using field-operating data collected from an in-service, containerized battery energy storage system deployed in Weihai, China. The studied battery cluster consists of 3,564 lithium-ion cells with \ch{LiFePO4}/graphite chemistry and a nominal capacity of 300 Ah per cell. As illustrated in Fig. \ref{fig3}a, the system is organized hierarchically, with cells assembled into modules, modules into packs, and packs connected in parallel at the cluster level. An open dataset covering eight months of continuous operation, from October 2024 onward, is constructed and released, encompassing more than 440 distinct charge and discharge operations, as summarized in Fig. \ref{fig3}b.

The dataset primarily contains operational measurements of voltage, current, temperature, and state of charge (SOC), representative of real-world BESS operation. Fig. \ref{fig3}c and \ref{fig3}d summarize the statistical distributions of operation duration and SOC variation, respectively. Most operations correspond to deep charging or discharging processes and persist for longer than 1.5 hours, providing sufficient temporal coverage for evaluating consistency-related behaviors under comparable operating conditions. These characteristics form the empirical basis for selecting representative standard operations in the construction of the inconsistency record dataset, as detailed in the Methods section.

Representative segments of the measured operational data are shown in Fig. \ref{fig3}e. In contrast to controlled laboratory experiments, the field data exhibit significantly higher complexity, including non-ideal current profiles, irregular load variations, and coupled thermal-electrical dynamics. Fig. \ref{fig3}f-h further illustrate the distributions of current, voltage, and temperature during different operations. While certain operations (e.g., Operations 3, 4, and 7) display relatively stable current profiles, others (e.g., Operation 6) exhibit pronounced fluctuations, reflecting the diverse grid-support roles of the BESS. Substantial voltage dispersion is observed in several battery packs, notably Packs 1, 8, and 9, while temperature distributions reveal heterogeneous spatial and temporal patterns across the cluster.

Collectively, these observations demonstrate that battery inconsistencies are pervasive under real-world BESS operation and evolve dynamically across operating scenarios, packs, and time scales. Importantly, such behaviors cannot be reliably characterized using static thresholds or isolated indicators alone. This highlights the necessity of an inconsistency-oriented, context-aware O\&M approach capable of interpreting heterogeneous operational data and supporting timely, decision-oriented maintenance actions under realistic operating conditions.

\subsection{Framework results for decision making}

\begin{figure}[h]
\centering
\includegraphics[page=4, width=\linewidth]{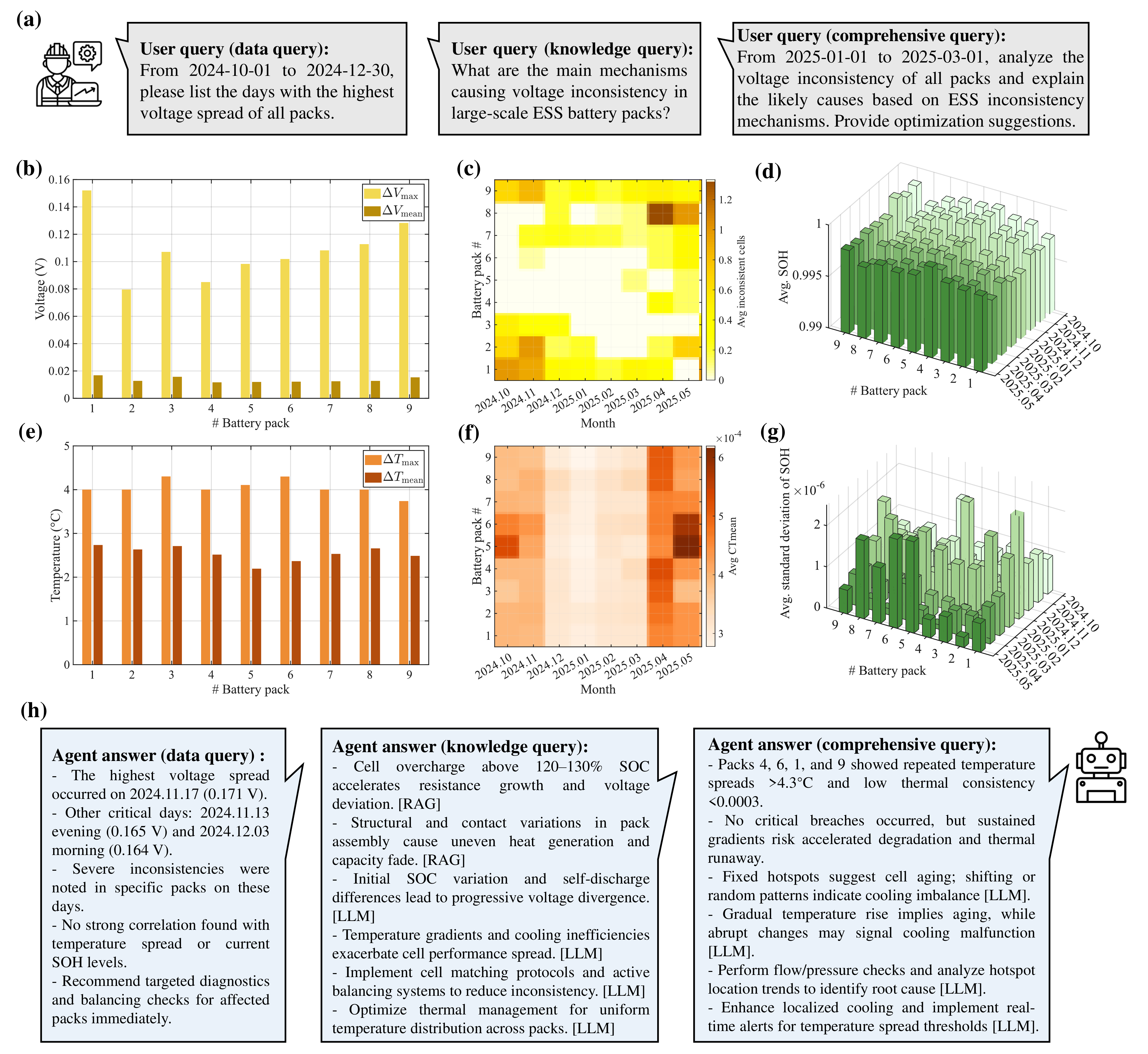}
\caption{Effectiveness of proposed explainable O\&M framework. \textbf{a} presents three sample queries from the O\&M engineer, including a data query, a knowledge query, and a comprehensive query. \textbf{b-g} display the inconsistency evaluation results from voltage, thermal and health perspectives. \textbf{b}, \textbf{e} show the $\Delta V_{\mathrm{max}}$, $\Delta V_{\mathrm{mean}}$, and $\Delta T_{\mathrm{max}}$, $\Delta T_{\mathrm{mean}}$ of all packs in Operation 1 in October 2024. \textbf{c}, \textbf{f} show the average inconsistent cell number and average TCC value of all packs during eight months. \textbf{d}, \textbf{g} show the average SOH estimation and average standard deviation of estimated SOH of all packs during eight months. \textbf{g} demonstrates the Agent answers for queries presented in \textbf{a} based on the inconsistency evaluation from the record dataset and the established knowledge dataset.}
\label{fig4}
\end{figure}

To evaluate the effectiveness of the proposed framework in supporting inconsistency-oriented O\&M of BESSs, a set of 30 representative user queries is designed in consultation with experienced O\&M engineers. The queries reflect common practical concerns encountered in real-world BESS operation and are categorized into three types: data-oriented, knowledge-oriented, and comprehensive queries that require joint interpretation of operational data and domain knowledge. Representative examples of each query type are illustrated in Fig. \ref{fig4}a, and the complete question set is provided in Supplementary Tables S1-S3.

Prior to query answering, representative standard operating conditions are selected from the full dataset to ensure that inconsistency evaluation results are comparable across operations, with results summarized in Supplementary Fig. S1. Based on these standard operations, multi-perspective inconsistency evaluation is performed from electrical, thermal, and health-related viewpoints, yielding structured indicators that form the record dataset. Figs. \ref{fig4}b-g summarize the resulting inconsistency characteristics across the eight-month operational period. From the electrical perspective, Fig. \ref{fig4}b shows that while the average voltage spread remains low for most packs, individual packs can exhibit pronounced maximum deviations driven by a small number of outlier cells. Temporal analysis further reveals that the number of inconsistent cells evolves dynamically over time rather than remaining constant (Fig. \ref{fig4}c). From the thermal perspective, Figs. \ref{fig4}e-f indicate that intra-pack temperature differences remain within a moderate range, but thermal consistency coefficients vary over time, reflecting gradual changes in thermal uniformity influenced by operating and environmental conditions. Health-related evaluation results (Fig. \ref{fig4}d and Fig. \ref{fig4}g) show that all packs remain in a healthy state during the study period, with stable and low-variance state-of-health estimates.

These structured inconsistency indicators provide the quantitative foundation for decision-making within the proposed framework. Using the constructed record dataset and the curated knowledge dataset (see Supplementary Fig. S3), the multi-agent system generates explainable responses to user queries in real time, as illustrated in Fig. \ref{fig4}h. For data-oriented queries, the framework accurately identifies critical operational events, such as days with the largest voltage dispersion, and summarizes relevant inconsistency patterns across packs. For example, in response to queries requesting periods of elevated voltage spread, the system not only reports the corresponding dates and magnitudes but also contextualizes these findings by correlating electrical indicators with thermal and health-related information, enabling targeted diagnostic follow-up.


For knowledge-oriented queries, the framework retrieves relevant domain knowledge from the knowledge dataset and integrates it with semantic reasoning to explain underlying inconsistency mechanisms. Retrieved information grounded in literature is explicitly distinguished from model-generated reasoning using provenance tags, such as \texttt{[RAG]} and \texttt{[LLM]}, ensuring transparency and traceability of the explanations. This design allows the system to deliver expert-level interpretations of inconsistency phenomena while maintaining clear attribution of information sources.


Comprehensive queries, which require both data-driven analysis and knowledge-based interpretation, highlight the full capability of the proposed framework. In these cases, the system first analyzes historical inconsistency patterns within the specified time window and then synthesizes relevant domain knowledge to explain likely causes and maintenance implications. By explicitly linking observed inconsistency behaviors with underlying physical mechanisms and operational conditions, the framework generates actionable, context-aware maintenance recommendations tailored to the specific BESS state.


In conventional BESS O\&M workflows, addressing such cross-dimensional questions typically requires manual inspection of multiple data streams and expert interpretation, often involving substantial time and labor costs. In contrast, the proposed framework enables interactive, explainable, and decision-oriented responses that directly connect inconsistency evaluation results with practical O\&M actions. These results demonstrate that inconsistency evaluation can be transformed from a diagnostic endpoint into an effective decision-support capability, substantially enhancing the efficiency, transparency, and scalability of real-world BESS O\&M.

\subsection{Performance testing}

\begin{figure}[h]
\centering
\includegraphics[page=5, width=\linewidth]{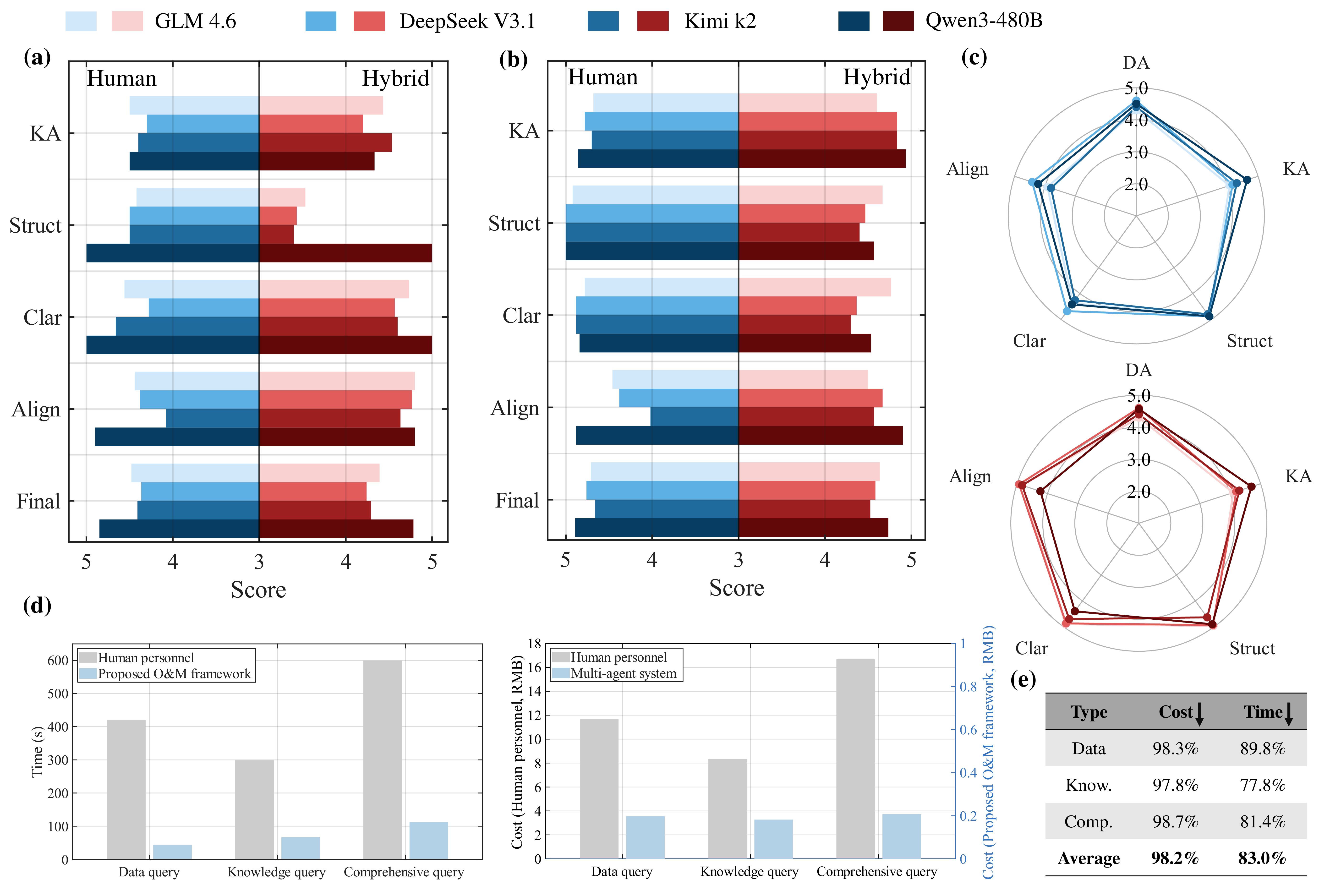}
\caption{\textbf{Evaluation of decisions made by the proposed O\&M framework and its cost-effectiveness (DA: data accuracy, KA: knowledge accuracy, Struct: structure quality, Clar: Clarity and brevity, Align: Alignment with user intent, Final: Final average score).} \textbf{a-c} Average scores from human experts and hybrid experts for the O\&M decision answers under different fully-parameterized models for data query,  knowledge query, and comprehensive query. \textbf{d} Comparison of the average response time and the average cost per query between the proposed method and human personnel. Response time is gathered based on a senior engineer from the studied BESS project; details are provided in Supplementary Note S1. \textbf{e} Average improvement on cost-effectiveness of the O\&M system compared to human personnel.}
\label{fig5}
\end{figure}

To evaluate the robustness and practical utility of the proposed framework for real-world BESS O\&M, its performance is assessed across four representative LLMs and multiple query types. The evaluation focuses on whether the framework can consistently deliver accurate, explainable, and decision-aligned responses under realistic operational scenarios, rather than on optimizing performance for any specific model implementation. Quantitative evaluation results are summarized in Fig. \ref{fig5}, with detailed model-level comparisons provided in Supplementary Note S2. A total of 30 representative queries are designed in consultation with experienced O\&M engineers, covering data-oriented, knowledge-oriented, and comprehensive queries that reflect common BESS O\&M tasks. Answer quality is assessed using a unified scoring framework that considers data accuracy, knowledge accuracy, structural quality, clarity and brevity, and alignment with user intent. Evaluations are conducted by five domain experts, complemented by hybrid assessments incorporating independent large-scale language models. Detailed scoring criteria, evaluation protocols, and evaluation settings are provided in the Supplementary Notes S1 and S2.


Across all evaluated models, the proposed framework consistently delivers high-quality responses for each query type (Fig. \ref{fig5}a–c). While minor variations are observed in structural organization and stylistic clarity, all tested models demonstrate strong capability in accurately interpreting inconsistency evaluation results and generating explainable, decision-relevant outputs. These results indicate that the effectiveness of the framework does not depend on a specific large language model, but instead arises from the underlying integration of structured operational records, curated domain knowledge, and task-oriented reasoning workflows.

Beyond response quality, the efficiency and economic implications of the proposed framework are quantitatively evaluated through comparison with conventional expert-driven O\&M practices. For the studied BESS, experienced engineers typically require several minutes to manually analyze operational data and respond to data-oriented, knowledge-oriented, and comprehensive queries, resulting in non-negligible labor costs per query. In contrast, the proposed framework enables near real-time query answering through cloud-based inference. As shown in Fig. \ref{fig5}d–e, the average response time is reduced by more than 80\%, while the per-query operational cost is reduced by approximately 98\% compared with manual expert analysis.




Such improvements directly address a key bottleneck in large-scale BESS deployment, where the scalability and cost of expert-driven O\&M increasingly constrain system reliability and economic performance. Together, these results demonstrate that the proposed framework is not only technically robust across different model implementations, but also economically viable and operationally scalable for deployment in real-world, large-scale energy storage systems.

\section{Discussion}\label{sec3}

This work introduces an inconsistency-driven, explainable O\&M paradigm for BESSs that bridges the long-standing gap between quantitative diagnostics and practical decision-making. By integrating structured inconsistency evaluation with knowledge-grounded semantic reasoning, the proposed framework transforms routine monitoring data into actionable, interpretable maintenance guidance. Validation using large-scale field data from an in-service BESS demonstrates that this approach can substantially reduce response time and operational cost while maintaining expert-level accuracy and transparency. A central contribution of this study is the reframing of battery inconsistency from a diagnostic endpoint into a decision-support signal. Existing BESS O\&M practices typically rely on numerical indicators or anomaly scores that require manual interpretation by domain experts. While effective for identifying abnormal behaviors, such approaches do not scale with the growing size and complexity of modern energy storage deployments. By contrast, the proposed framework explicitly links observed inconsistency patterns with underlying physical mechanisms and operational implications, enabling timely and context-aware O\&M actions. This shift from metric-centric monitoring to decision-oriented interpretation represents a fundamental change in how inconsistency information is utilized in real-world BESS operation.


The explainability of the proposed framework is grounded in its dual-dataset design. The record dataset provides a compact and interpretable representation of daily operational states by summarizing electrical, thermal, and aging-related inconsistencies under standardized operating conditions. In parallel, the knowledge dataset distills domain expertise from the literature into structured, retrievable knowledge units. By constraining semantic reasoning to these two curated datasets, the framework ensures that generated responses remain transparent, traceable, and verifiable. Importantly, LLMs are not employed as black-box predictors, but as semantic reasoning components that translate structured data and expert knowledge into human-interpretable explanations and maintenance recommendations.

From an energy-systems perspective, the proposed framework directly addresses key challenges associated with the large-scale deployment of BESSs. As energy storage systems increasingly participate in grid balancing, peak shaving, and renewable energy integration, delays or ambiguities in maintenance decision-making can propagate into reliability risks and economic losses. By enabling near real-time, explainable interpretation of inconsistency behaviors, the framework supports more reliable and cost-effective participation of BESSs in modern power systems. The demonstrated reductions in response time and O\&M cost highlight the potential of decision-oriented intelligence to alleviate one of the major non-technical barriers to scaling energy storage infrastructure.


Beyond the specific application to battery inconsistency, the proposed framework is inherently extensible. Its model-agnostic and system-independent design allows it to accommodate diverse analytical algorithms and heterogeneous data sources without modifying the core decision-making logic. Although demonstrated at the battery-pack level, the framework can naturally extend across cell, module, cabinet, and system levels to support degradation analysis, fault precursor identification, and safety monitoring. More broadly, the underlying design philosophy—combining structured operational records with knowledge-grounded semantic reasoning—is applicable to other complex energy infrastructures, including power grids, industrial energy systems, and cyber-physical assets that require explainable and scalable O\&M decision support.


Several limitations of the current study warrant further investigation. First, the size of the knowledge dataset is constrained by the limited number of curated literature sources. Expanding this dataset to incorporate additional high-quality resources, such as international standards, technical guidelines, and operational best practices, would further enhance the completeness and robustness of knowledge-based reasoning. Second, data-intensive queries spanning extended operational horizons may result in excessively long input prompts, as large volumes of historical records are aggregated for a single inference. This can significantly increase token usage and lead to computational overhead, potentially degrading system efficiency. More efficient data abstraction and partitioning strategies will be essential for long-term, multi-year deployments.


Future work will focus on addressing these limitations and extending the framework’s capabilities. Potential directions include the incorporation of more adaptive routing strategies to refine query interpretation, the integration of function-calling mechanisms to preprocess large-scale operational data before semantic reasoning, and the adoption of multimodal models to support visualization, reporting, and interactive analytics. Together, these developments will further strengthen the proposed framework as a scalable and reusable foundation for intelligent, explainable O\&M in data-intensive energy systems.


\section{Methods}\label{sec4}
\subsection{Selection of standard operation}
In practical BESS operations, current profiles rarely follow an ideal Constant Current (CC) pattern and instead exhibit varying degrees of fluctuation. To construct a record dataset in which such inconsistencies can be evaluated in a comparable and meaningful manner, it is necessary to first identify a subset of representative standard operations from all recorded operations. This is achieved by screening operations based on their duration and current stability, so as to retain sufficiently long and quasi-steady operating segments while excluding operations with pronounced fluctuations. In this way, a consistent reference set of standard operations is established, which serves as the basis for subsequent inconsistency evaluation and record dataset construction. The details are illustrated in Supplementary Note S3.

\subsection{Construction of record dataset}

The record dataset is constructed by integrating basic operating information with summarized indicators characterizing electrical, thermal, and aging-related inconsistencies of battery packs under standard operating conditions. For each operation, basic metadata including the operation date, start and end time, and operation type (charging or discharging) are first recorded.

Subsequently, the inconsistency characteristics are organized into three matrices corresponding to voltage, thermal, and health perspectives. The voltage inconsistency matrix is defined as:
\begin{equation}
\mathbf{V} =
\begin{bmatrix}
\Delta V_{\mathrm{max},1} & \Delta V_{\mathrm{max},2} & \cdots & \Delta V_{\mathrm{max},9} \\
\Delta V_{\mathrm{mean},1} & \Delta V_{\mathrm{mean},2} & \cdots & \Delta V_{\mathrm{mean},9} \\
\mathrm{num}_1 & \mathrm{num}_2 & \cdots & \mathrm{num}_9
\end{bmatrix},
\end{equation}
where $\Delta V_{\mathrm{max},i}$ and $\Delta V_{\mathrm{mean},i}$ denote the maximum and average cell-voltage ranges of Pack~$i$, respectively, reflecting the severity and overall level of electrical imbalance, and $\mathrm{num}_i$ represents the number of cells identified as electrically inconsistent in Pack $i$.

Similarly, the thermal inconsistency matrix is constructed as:
\begin{equation}
\mathbf{T} =
\begin{bmatrix}
\Delta T_{\mathrm{max},1} & \Delta T_{\mathrm{max},2} & \cdots & \Delta T_{\mathrm{max},9} \\
\Delta T_{\mathrm{mean},1} & \Delta T_{\mathrm{mean},2} & \cdots & \Delta T_{\mathrm{mean},9} \\
\mathrm{TCC}_1 & \mathrm{TCC}_2 & \cdots & \mathrm{TCC}_9
\end{bmatrix},
\end{equation}
where $\Delta T_{\mathrm{max},i}$ and $\Delta T_{\mathrm{mean},i}$ represent the maximum and average temperature differences of Pack~$i$, respectively, and $\mathrm{TCC}_i$ denotes the thermal consistency coefficient, which captures persistent thermal imbalance within the pack.

From the aging perspective, the health condition of the battery packs is summarized by the health matrix:
\begin{equation}
\mathbf{H} =
\begin{bmatrix}
\mathrm{SOH}_1 & \mathrm{SOH}_2 & \cdots & \mathrm{SOH}_9
\end{bmatrix},
\end{equation}
where $\mathrm{SOH}_i$ represents the state of health of Pack~$i$, reflecting its capacity degradation status. The derivation of $\Delta V_{\mathrm{max},i}$,  $\Delta V_{\mathrm{mean},i}$, $\mathrm{num}_i$,  $\Delta T_{\mathrm{max},i}$, $\Delta T_{\mathrm{mean},i}$, $\mathrm{TCC}_i$ and $\mathrm{SOH}_i$ can be refereed to Supplementary Note S4 for detail.

All the above information is consolidated into a single entry in the record dataset, representing one standard operation. Each entry uses the operation date as its key to facilitate indexing and retrieval.

\subsection{Construction of knowledge dataset}
To enhance the professionalism and domain accuracy of the content generated by the proposed framework, a dedicated knowledge dataset is constructed to support retrieval-augmented generation (RAG). The dataset is built through a three-stage pipeline consisting of literature selection, knowledge distillation, and dataset construction.

First, representative literature is collected from target domains, including BESS applications, battery inconsistency mechanisms, and battery fault mechanisms. Review articles are prioritized as they provide comprehensive and structured domain knowledge. In this study, ten carefully selected review articles are used as foundational sources.
Next, an LLM-driven knowledge distillation strategy is employed to extract structured and semantically coherent knowledge units from the selected literature, avoiding the limitations of conventional fixed-length text segmentation. This step produces concise knowledge slices that preserve conceptual integrity and domain relevance. 
Finally, each knowledge slice is associated with a compact semantic key that represents its core meaning. After manual verification, these semantic keys are embedded and stored in the vector database. During retrieval, user queries are matched directly against the semantic keys, enabling efficient and accurate knowledge retrieval.

Through this pipeline, the constructed knowledge dataset provides a reliable foundation for expert-level reasoning in the proposed multi-agent intelligent system. Further details are presented in the Supplementary Note 5.

\subsection{Multi-agent system for intelligent O\&M}
The proposed multi-agent system integrates a structured record dataset describing battery-pack inconsistencies with a curated knowledge dataset containing expert-level information on BESS applications. Together, they support an online query-answering service that enables efficient, explainable, and knowledge-grounded operation and maintenance (O\&M) for BESS engineers. As illustrated in Fig. \ref{fig2}(b3), the system follows a modular workflow in which user queries are first analyzed and then routed to specialized processing pipelines according to their intent.

For data queries, which focus on extracting or analyzing historical operational records, the system emphasizes structured data access and quantitative reasoning. After identifying the temporal scope of the query, relevant records are retrieved through date-based indexing. Data synthesis and analysis are then performed to generate responses that are strictly grounded in the retrieved records, ensuring factual consistency and traceability. This route is designed for tasks such as trend inspection, comparison across operations, and anomaly summarization.

For knowledge queries, which seek expert understanding of BESS mechanisms, faults, or mitigation strategies, the system adopts a retrieval-augmented generation (RAG) paradigm. External knowledge entries are retrieved from the domain knowledge dataset and combined with model-based expert reasoning to generate grounded explanations. Depending on the relevance of retrieved evidence, the response dynamically balances retrieved knowledge and internal reasoning, enabling both factual accuracy and interpretability.

For comprehensive queries, which require joint interpretation of operational data and domain knowledge, the system coordinates multiple agents to decompose the query, process data-driven and knowledge-driven components in parallel, and synthesize the results into a unified response. This route enables the system to provide actionable insights that connect observed data patterns with underlying physical mechanisms and engineering implications. The detailed agent interactions, prompt designs, and implementation strategies for each routing path are provided in the Supplementary Note S6.

\section*{Data availability}
The data is still under refinement and will be fully organized and uploaded to the GitHub repository (https://github.com/Semingwk/From-inconsistency-to-decision-explainable-O-M-of-battery-energy-storage-syst) upon publication.

\section*{Code availability}
The code is still under refinement and will be fully organized and uploaded to the GitHub repository (https://github.com/Semingwk/From-inconsistency-to-decision-explainable-O-M-of-battery-energy-storage-syst) upon publication.

\section*{Supplementary information}
The supplementary information is available as an individual file attached during submission.

\bibliography{sn-bibliography}

\end{document}


\renewcommand{\figurename}{Fig.}
\renewcommand{\thefigure}{S\arabic{figure}}
\renewcommand{\thetable}{S\arabic{table}}

\noindent{\Large\bfseries Supplementary Information:}
    
\vspace{1em}

\parbox{1\textwidth}{%
  \centering
  \bfseries
  \fontsize{18}{20}\selectfont   
  From inconsistency to decision: explainable operation and maintenance of battery energy storage systems
}
\vspace{1em}
\begin{center}
    {\large
    Jingbo Qu$^{1,4,5,\dagger}$,
    Yijie Wang$^{2,3,\dagger,*}$,
    Yujie Fu$^{2}$,
    Putai Zhang$^{2}$,
    Weihan Li$^{4,5,*}$,
    Mian Li$^{2,3,*}$
    }
\end{center}

\vspace{1em}

\noindent
$^{1}$ Global College, Shanghai Jiao Tong University, Shanghai 200240, China

\noindent
$^{2}$ Global Institute of Future Technology, Shanghai Jiao Tong University, Shanghai 200240, China

\noindent
$^{3}$ Key Laboratory of Urban Complex Risk Control and Resilience Governance,  
Shanghai Jiao Tong University, Shanghai 200030, China

\noindent
$^{4}$ Center for Ageing, Reliability and Lifetime Prediction of Electrochemical and Power Electronic Systems (CARL),  
RWTH Aachen University, Aachen 52074, Germany

\noindent
$^{5}$ Institute for Power Electronics and Electrical Drives (ISEA),  
RWTH Aachen University, Aachen 52074, Germany

\vspace{1em}

\noindent
$\dagger$ These authors contributed equally to this work.

\noindent
$^{*}$ Corresponding author:  
\texttt{yijiewang@sjtu.edu.cn},  
\texttt{Weihan.Li@isea.rwth-aachen.de},
\texttt{mianli@sjtu.edu.cn}
\newpage

\section*{Supplementary Figure}

\begin{figure}[h]
\centering
\includegraphics[width=\linewidth]{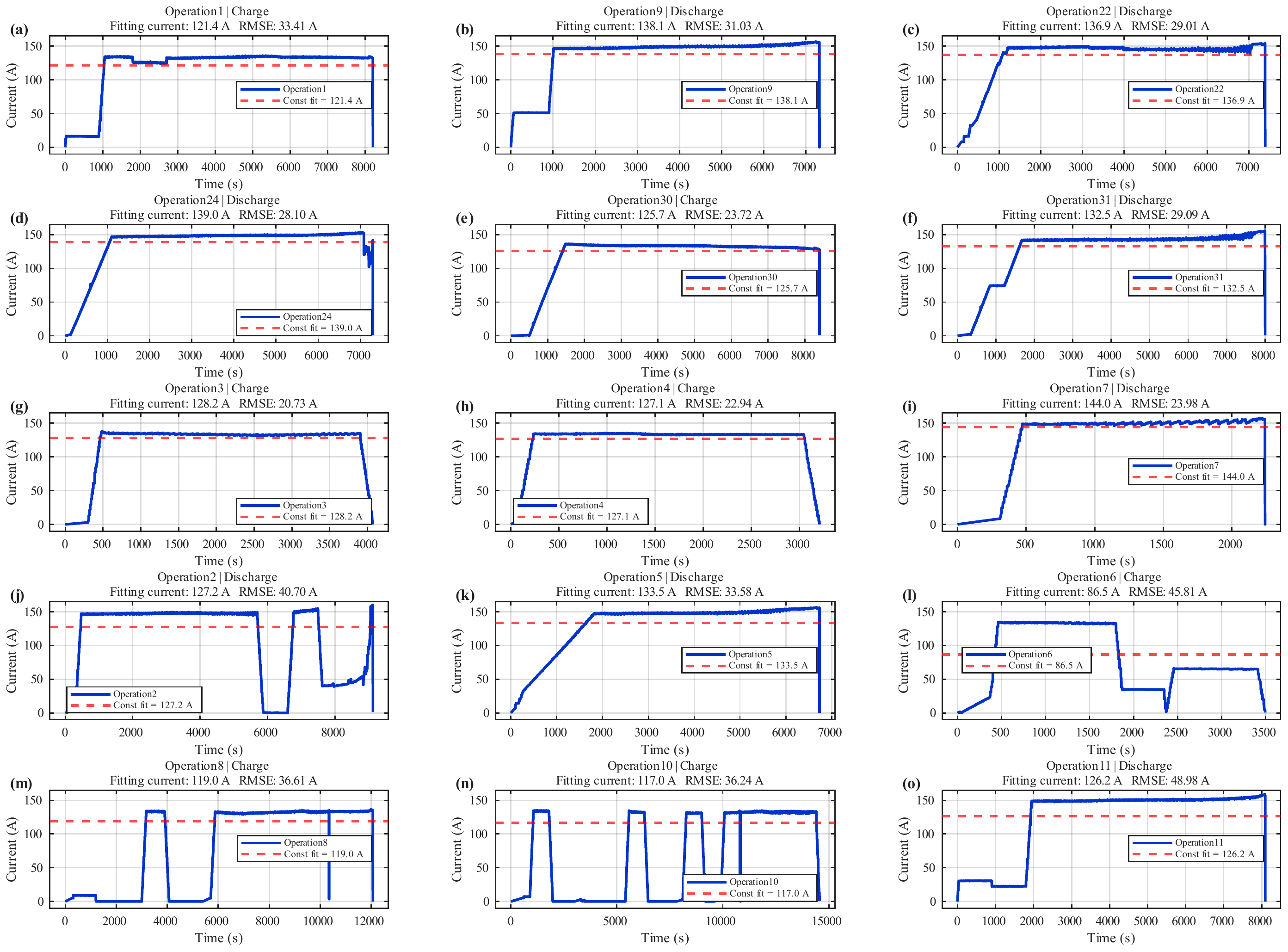}
\caption{Results of standard operation selection in October 2024. Both the fitted current and the corresponding fitting error are shown, demonstrating that the proposed selection method can successfully identify operations with stable current profiles, even under the complex and dynamic loading conditions typical of BESS applications. (a)-(f): The selected standard operations. (g)-(i): Operations that are not selected due to short duration time. (j)-(o): Operations that are not selected due to a large fitting error resulting from fluctuations.}
\label{figs1}
\end{figure}

\newpage

\begin{figure}[h]
\centering
\includegraphics[width=\linewidth]{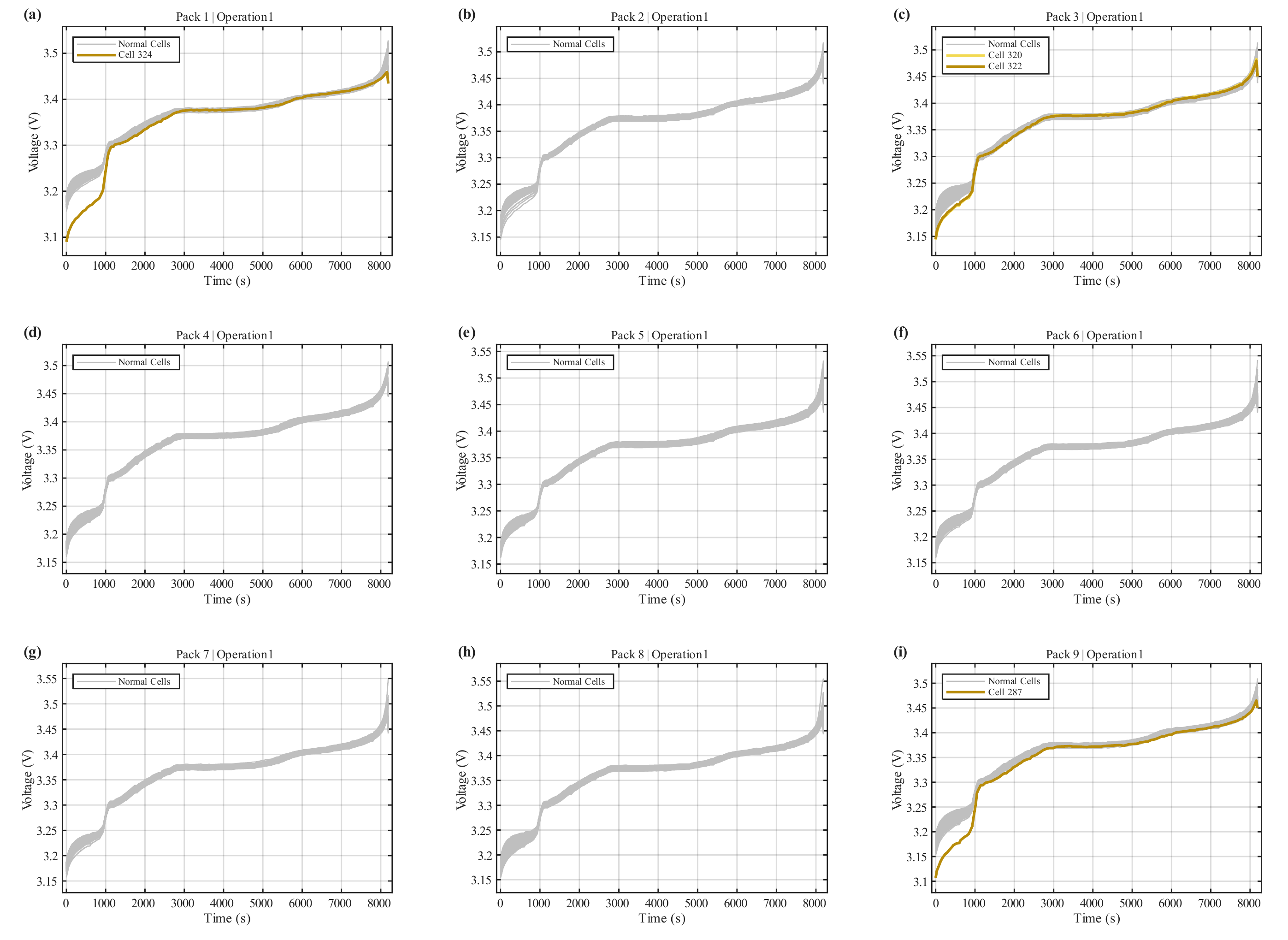}
\caption{Voltage inconsistency identification results for all battery packs in Operation 1 in October 2024. Pack 1, 2 and Pack 9 exhibit abnormal outlier cells with relatively large deviations, whereas the remaining packs maintain good overall consistency}
\label{figs2}
\end{figure}

\newpage

\begin{figure}[h]
\centering
\includegraphics[width=\linewidth]{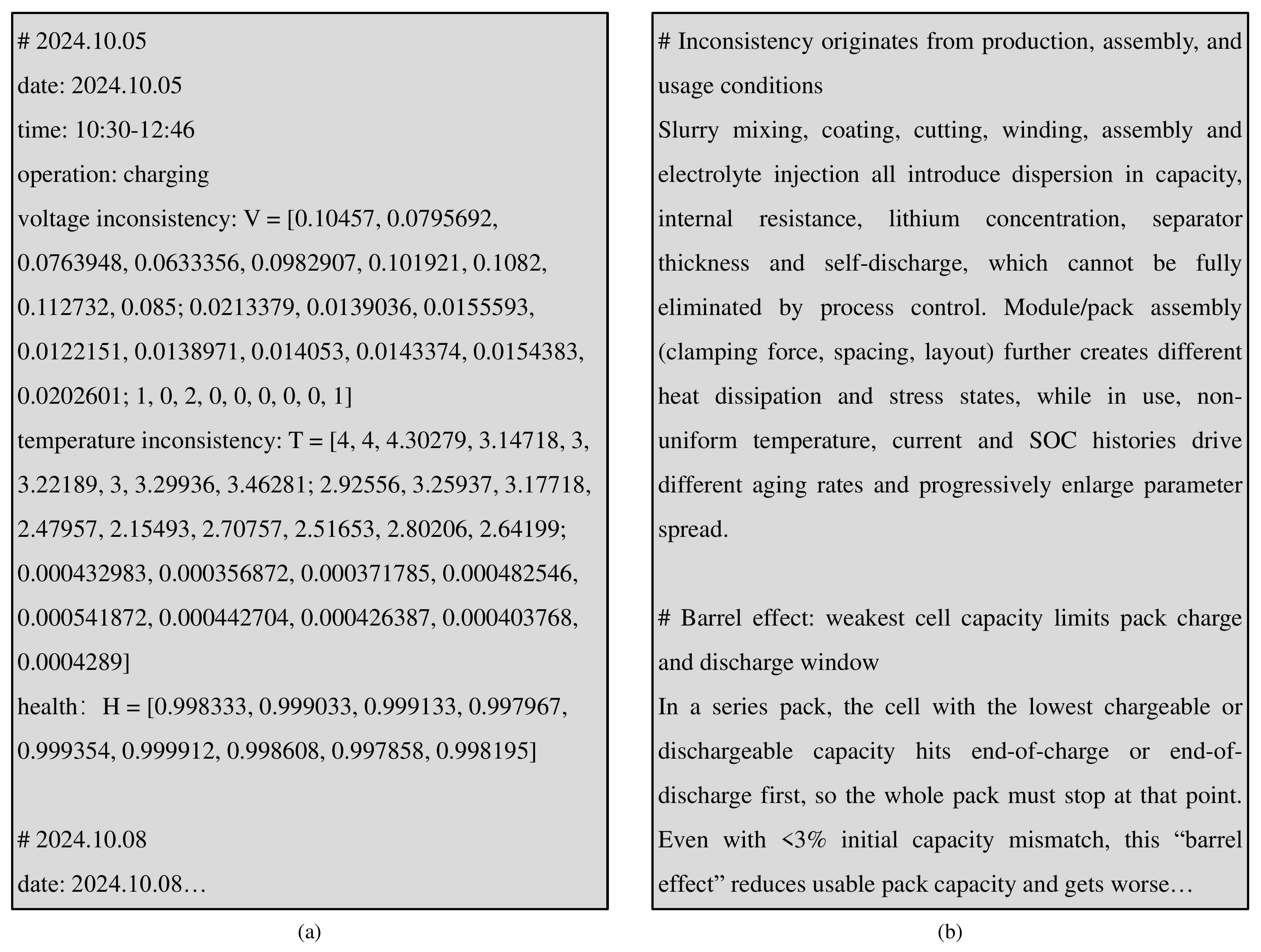}
\caption{Sample segments for the record dataset and knowledge dataset. (a) Sample chunked segments of the record dataset. (b) Sample chunked segments of the knowledge dataset.}
\label{figs3}
\end{figure}

\newpage

\begin{figure}[h]
\centering
\includegraphics[width=\linewidth]{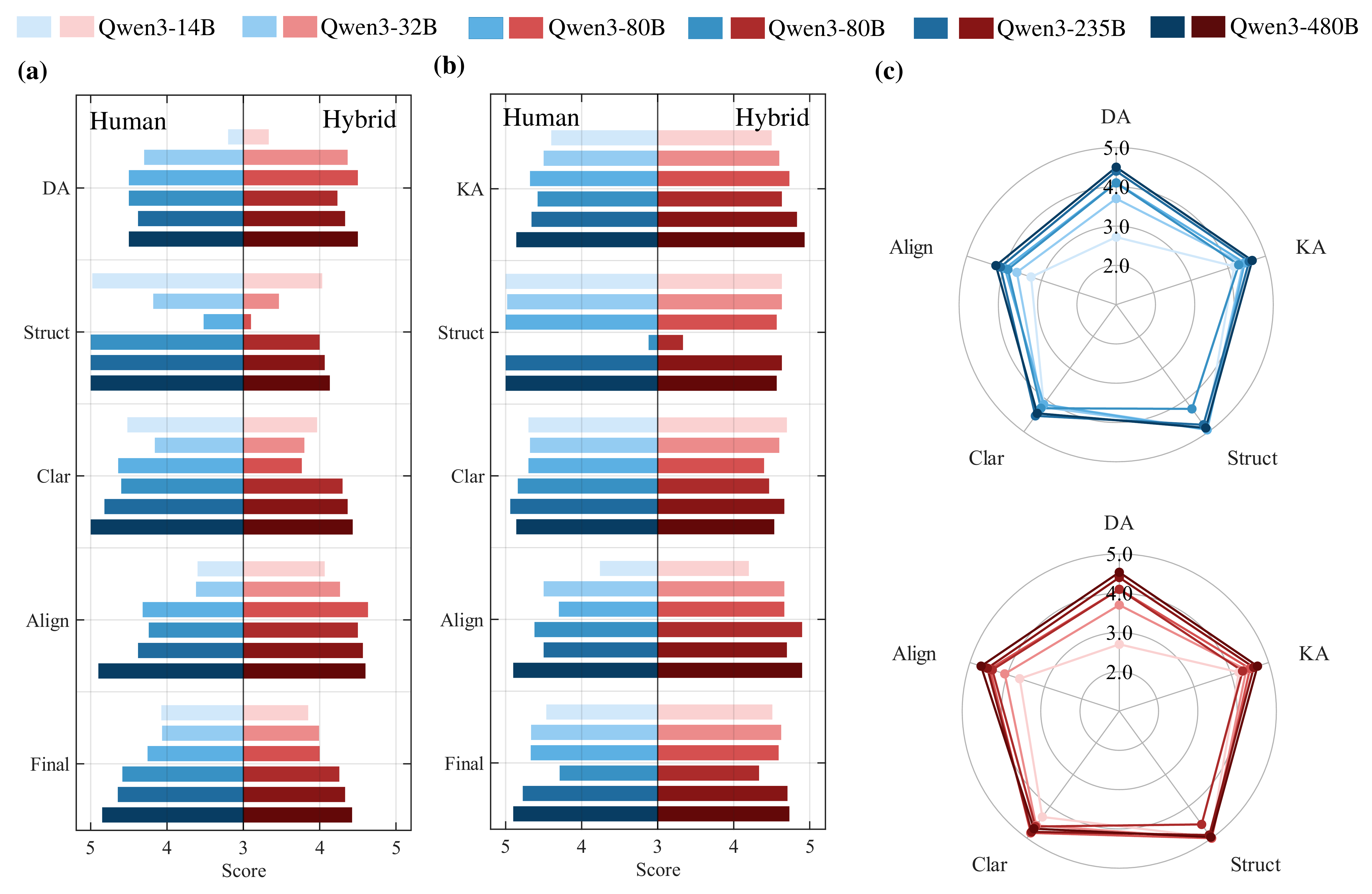}
\caption{Evaluation results from human and hybrid experts for the proposed framework under varying LLM parameter scales. (DA: data accuracy, KA: knowledge accuracy, Struct: structure quality, Clar: Clarity and brevity, Align: Alignment with user intent, Final: Final average score). (a) Average scores from human experts and hybrid experts for data query. (b) Average scores from human experts and hybrid experts for knowledge query. (c) Average scores from human experts and hybrid experts for comprehensive query.}
\label{figs4}
\end{figure}

\newpage

\section*{Supplementary Note}
\subsection*{Note S1: Scoring criteria, evaluation protocols, and evaluation setting}

The evaluation of answer quality is based on five criteria, described as follows:

\begin{enumerate}
\item \textbf{Data accuracy}: Ensures that the LLM retrieves and analyzes information strictly from the record dataset without hallucination.
\item \textbf{Knowledge accuracy}: Ensures that the LLM correctly explains underlying mechanisms using both retrieved knowledge (RAG results) and its internal domain understanding.
\item \textbf{Structure quality}: Ensures that the LLM's response follows the required concise and well-organized format.
\item \textbf{Clarity and brevity}: Ensures that the LLM produces responses that are direct, clear, and immediately understandable to operational personnel, while avoiding redundancy, repeated concepts, and vague expressions.
\item \textbf{Alignment with user intent}: Ensures that the LLM's final answer accurately corresponds to the user's intent and directly addresses the user's request.
\end{enumerate}

Each evaluation criterion has a maximum score of 5 and a minimum score of 0. For data queries, the knowledge accuracy criterion is excluded, and for knowledge queries, the data accuracy criterion is excluded, resulting in a total possible score of 20 for each of these categories. Comprehensive queries consider all five criteria, yielding a total possible score of 25. The detailed scoring rules are provided as followings.
\vspace{1em}

\noindent\textbf{Scoring scale:} 0-5 points per dimension (integer values only)

\noindent\textbf{Applicable routing path:}
\begin{itemize}
    \item Data query routing: Data accuracy, structure quality, clarity and brevity, alignment with user intent are evaluated.
    \item Knowledge query routing: Knowledge accuracy, structure quality, clarity and brevity, alignment with user intent are evaluated.
    \item Comprehensive query routing: Data accuracy, knowledge accuracy, structure quality, clarity and brevity, alignment with user intent are evaluated.
\end{itemize}

\vspace{0.5cm}

\noindent\textbf{Data accuracy}

\textbf{Applicable to:} Data query routing and Comprehensive query routing

\textbf{Evaluation focus:}
\begin{itemize}
    \item Correct interpretation of V/T/H matrices from the record dataset
    \item Accuracy of cited numerical values (dates, matrix values, trends)
    \item No hallucination of abnormal patterns or fabricated data
    \item All data-related statements must be grounded in retrieved records
\end{itemize}

\begin{table}[h]
\centering
\small
\begin{tabular}{cp{12cm}}
\toprule
\textbf{Score} & \textbf{Description} \\
\midrule
\textbf{5} & All data references are completely accurate. V/T/H matrices correctly interpreted according to their physical meanings (e.g., V\_row1 = worst-case voltage spread, T\_row3 = thermal consistency coefficient). No hallucinations. Numerical values match retrieved records exactly. \\
\addlinespace
\textbf{4} & Data mostly accurate with 1-2 minor numerical interpretation deviations (e.g., slightly misreading a trend direction) that do not affect overall conclusions. \\
\addlinespace
\textbf{3} & Data largely accurate but contains obvious numerical misreading or misunderstanding of matrix semantics (e.g., confusing V\_row1 with V\_row2). \\
\addlinespace
\textbf{2} & Multiple data errors present, or cites data not found in retrieved records. May include fabricated anomalies or pack behaviors. \\
\addlinespace
\textbf{1} & Extensive data hallucination. Severe misinterpretation of matrix meanings. Most numerical claims unverifiable. \\
\addlinespace
\textbf{0} & Completely fabricated data, OR refuses to perform analysis citing ``insufficient data'' when matrices are actually available. \\
\bottomrule
\end{tabular}
\end{table}

\newpage

\noindent\textbf{Knowledge accuracy}

\textbf{Applicable to:} Knowledge query routing and Comprehensive query routing

\textbf{Evaluation focus:}
\begin{itemize}
    \item Correct explanation of BESS mechanisms 
    \item Engineering reasoning is sound and aligned with the BESS domain knowledge
    \item No fabrication of non-existent BESS mechanisms
\end{itemize}

\begin{table}[h]
\centering
\small
\begin{tabular}{cp{12cm}}
\toprule
\textbf{Score} & \textbf{Description} \\
\midrule
\textbf{5} & All BESS mechanism explanations are completely correct (e.g., thermal propagation, cooling system impacts, SOC/SOH balancing). Engineering reasoning is sound and practical. \\
\addlinespace
\textbf{4} & Knowledge mostly correct with 1-2 minor imprecisions in mechanism explanation (e.g., slightly oversimplifying a thermal management concept) that do not affect actionability. \\
\addlinespace
\textbf{3} & Knowledge generally correct but contains obvious mechanism misunderstandings or engineering reasoning that conflicts with BESS best practices. \\
\addlinespace
\textbf{2} & Multiple knowledge errors, or severe misinterpretation of RAG content. May include hallucinated BESS mechanisms. \\
\addlinespace
\textbf{1} & Extensive knowledge hallucination. Fabricates non-existent BESS mechanisms or safety standards. \\
\addlinespace
\textbf{0} & Completely incorrect knowledge, or fails to provide any knowledge-based explanation. \\
\bottomrule
\end{tabular}
\end{table}

\noindent\textbf{Structure quality}

\textbf{Applicable to:} All routing paths

\textbf{Evaluation focus:}
\begin{itemize}
    \item Correct number of bullet points (3-6)
    \item Each bullet is a single sentence under 25 words
    \item Proper labeling: \texttt{[Data]}, \texttt{[Knowledge]}, \texttt{[Integrated]} prefixes for comprehensive query routing
    \item Proper source tags: \texttt{[RAG]}, \texttt{[LLM]}, or \texttt{[RAG][LLM]} at end of each bullet
\end{itemize}

\begin{table}[h]
\centering
\small
\begin{tabular}{cp{12cm}}
\toprule
\textbf{Score} & \textbf{Description} \\
\midrule
\textbf{5} & Strictly adheres to 3-6 bullets. Every bullet is a single sentence ($<$25 words). All bullets correctly labeled with routing-appropriate prefixes (\texttt{[Data]}/\texttt{[Knowledge]}/\texttt{[Integrated]}) and source tags (\texttt{[RAG]}/\texttt{[LLM]}). \\
\addlinespace
\textbf{4} & Correct bullet count (3-6). 1-2 bullets slightly exceed word limit (26-30 words). Format mostly compliant with minor issues (e.g., one missing \texttt{-~} prefix). Labels are generally correct. \\
\addlinespace
\textbf{3} & Correct bullet count but multiple bullets significantly exceed word limit (31-40 words), OR formatting has obvious issues (e.g., missing prefixes, improper nesting). Label usage is inconsistent. \\
\addlinespace
\textbf{2} & Bullet count incorrect (fewer than 3 or more than 6), AND/OR poor word control (many bullets $>$40 words), AND/OR chaotic formatting (mixed lists/paragraphs). Labels are missing or misused. \\
\addlinespace
\textbf{1} & Not in bullet format at all (e.g., paragraph blocks), OR severely chaotic structure. \\
\addlinespace
\textbf{0} & No structured output provided. \\
\bottomrule
\end{tabular}
\end{table}

\noindent\textbf{Clarity and brevity}

\textbf{Applicable to:} All routing paths

\textbf{Evaluation focus:}
\begin{itemize}
    \item Each bullet is direct, powerful, and immediately understandable by O\&M staff
    \item Avoids redundancy, concept repetition, and ambiguous expressions
    \item Technical terms are used appropriately (not oversimplified or overcomplicated)
    \item No circular reasoning or verbose explanations
\end{itemize}

\begin{table}[h]
\centering
\small
\begin{tabular}{cp{12cm}}
\toprule
\textbf{Score} & \textbf{Description} \\
\midrule
\textbf{5} & Every bullet is direct, powerful, and immediately actionable for O\&M staff. Zero redundancy. No ambiguity. Technical precision balanced with accessibility. \\
\addlinespace
\textbf{4} & Clear expression with 1-2 bullets slightly verbose or containing minor redundancy (e.g., repeating a concept already stated). \\
\addlinespace
\textbf{3} & Mostly clear but contains obvious concept repetition across bullets, OR uses vague expressions (e.g., ``may be related to various factors''). \\
\addlinespace
\textbf{2} & Multiple clarity issues: significant redundancy, ambiguous statements, or overly complex jargon that obscures meaning. \\
\addlinespace
\textbf{1} & Confusing and hard to understand. Circular reasoning. Excessive technical jargon without explanation. \\
\addlinespace
\textbf{0} & Completely incomprehensible. \\
\bottomrule
\end{tabular}
\end{table}

\newpage

\noindent\textbf{Alignment with user intent}

\textbf{Applicable to:} All routing paths

\textbf{Evaluation focus:}
\begin{itemize}
    \item Correctly selects the appropriate routing path
    \item Actually answers the user's core question
    \item Provides actionable recommendations when requested
    \item Avoids off-topic content
\end{itemize}

\begin{table}[h]
\centering
\small
\begin{tabular}{cp{12cm}}
\toprule
\textbf{Score} & \textbf{Description} \\
\midrule
\textbf{5} & Perfectly answers the user's core question. Routing path selection is correct. Provides concrete, actionable suggestions directly addressing user needs. Zero off-topic content. \\
\addlinespace
\textbf{4} & Substantially answers the core question. Routing appropriate. 1-2 points slightly deviate from focus or lack depth in addressing user needs. \\
\addlinespace
\textbf{3} & Partially answers the question but has obvious omissions or deviates from key points. May include some irrelevant content. \\
\addlinespace
\textbf{2} & Routing path selection is wrong. Mostly off-topic or fails to address the user's primary concern. Recommendations (if any) are vague or not actionable. \\
\addlinespace
\textbf{1} & Severely off-topic. Does not answer the core question. \\
\addlinespace
\textbf{0} & Completely unresponsive to user needs. \\
\bottomrule
\end{tabular}
\end{table}

The proposed framework is evaluated using two complementary scoring approaches: human expert evaluation and hybrid expert evaluation. Specifically, the human expert evaluation is conducted by five domain experts specializing in Battery Energy Storage Systems (BESS). The hybrid evaluation combines assessments from both human experts and three flagship large language models (LLMs), namely GPT-5.1 \cite{gpt5}, Claude Sonnet 4.5 \cite{claude_sonnet_45}, and Gemini 3 Pro \cite{gemini_3_pro}.

Given that the record dataset and the knowledge dataset provide ground-truth answers, human experts are responsible for scoring data accuracy and knowledge accuracy to ensure correctness and reliability. In contrast, the LLMs primarily evaluate criteria such as structure, clarity, and alignment, which are less dependent on domain-specific ground truth. Detailed information regarding expert participation is provided as follows.

The expert scoring was performed by a panel of senior engineers with extensive expertise in energy storage systems. All experts either hold doctoral degrees or possess substantial industrial experience as master's or bachelor's degree holders with long-term professional practice. They have been continuously engaged in frontline engineering work and have accumulated practical experience in the development, testing, and quality management of BESS. Their expertise spans key domains including Battery Management Systems (BMS), energy storage system architecture, power station development, structural design, quality engineering, and battery testing.

In this study, the first five experts listed in Table below participated in the full evaluation and provided complete scores across all criteria. The remaining three experts contributed specifically to the evaluation of data accuracy and knowledge accuracy. To ensure consistency, reliability, and credibility of the expert-based evaluation, all participants strictly followed the predefined scoring rubric and evaluation standards.

In addition, Shi Eng.\ was invited to answer all queries manually based on the provided record and knowledge datasets. The average response time for each type of routing query was recorded and shown in Fig. 5d.

\begin{table}[htbp]
\centering
\label{tab:experts}
\renewcommand{\arraystretch}{1.2}
\begin{tabular}{cccc}
\hline
\textbf{Expert} & \textbf{Degree} & \textbf{Position} & \textbf{Experience (year)} \\
\hline
Jiang Eng. & M.Sc. & Early-stage Quality Engineer & 8+ \\
Han Eng. & Ph.D. & BMS System Engineer & 4+ \\
Shi Eng. & M.Sc. & BMS System Engineer & 6+\\
Li Eng. & B.Eng. & BMS System Engineer & 10+  \\
Chen Eng. & B.Eng. & TMS System Engineer & 7+  \\
Zhang Eng. & Ph.D. & Energy Storage System Architect & 5+  \\
Yao Eng. & M.Sc. & Battery Testing and Validation Engineer & 4+  \\
Wang Eng. & M.Sc. & Battery Testing and Validation Engineer & 6+  \\
\hline
\end{tabular}
\end{table}

\newpage

\subsection*{Note S2: Evaluation setup and Comparisons on LLMs}

To assess the robustness and deployability of the proposed framework, we conduct evaluations using both fully parameterized large-scale LLMs and models of varying parameter scales. This design aims to verify that the framework maintains consistent behavior across heterogeneous model capacities, rather than being tailored to a specific model configuration.

Specifically, several fully parameterized LLMs, including GLM 4.6, DeepSeek V3.1, Kimi-k2-turbo-preview (denoted as Kimi-k2), and Qwen3-480B, are employed for comparison. In addition, models spanning a wide range of parameter scales—from 8B to 480B—are evaluated to examine scalability and deployment flexibility. These include Qwen3-8B, Qwen3-14B, Qwen3-32B, Qwen3-Next-80B-A3B-Instruct (denoted as Qwen3-80B), Qwen3-235B-A22B, and Qwen3-Coder-480B-A35B-Instruct.
For consistency, all agents within the multi-agent system are instantiated using the same model scale in each experiment. The comparative evaluation results across different model configurations are summarized in Fig. 5 and Fig. \ref{figs4}.

Answer quality is assessed using a unified scoring framework based on five criteria mentioned in Note S1. Both human expert evaluation and hybrid evaluation are adopted according to Note S1.

For data-oriented queries, as shown in Fig. 5a, all models exhibit comparable performance in data accuracy, while Qwen3-480B achieves the highest overall score (4.85), primarily due to superior structural quality, clarity, and intent alignment. In contrast, GLM 4.6, DeepSeek V3.1, and Kimi k2 demonstrate relatively weaker structural organization, suggesting limited sensitivity to prompt constraints on response formatting. For knowledge-oriented queries (Fig. 5b), all models perform consistently well across evaluation dimensions, with Qwen3-480B receiving the highest score in intent alignment. Overall, all models achieve scores over 4, indicating strong capability in knowledge retrieval and explanation.
For comprehensive queries, human expert evaluation shows that DeepSeek V3.1 attains the highest scores in clarity and intent alignment, while Qwen3-480B achieves the highest knowledge accuracy. Structural quality and data accuracy remain consistently high across all models. Under hybrid evaluation, Qwen3-480B exhibits slightly lower scores in clarity and alignment, yet continues to outperform others in knowledge accuracy, whereas data accuracy and structural quality remain comparable among the evaluated models.

Comparisons on LLMs with different parameter scales are provided in Fig. \ref{figs4} for reference. For data-query tasks (Fig. \ref{figs4}(a)), most models achieve high scores in data accuracy, with the exception of the smallest-scale variant. Structure quality exhibits greater variability across model scales: mid-sized models show noticeably lower scores, whereas both smaller and larger models produce more stable and well-formatted outputs. The final average scores follow a clear upward trend with increasing parameter size. For knowledge-query tasks (Fig. \ref{figs4}(b)), a pronounced drop in structure quality is observed for the mid-sized model, despite its strong performance in knowledge accuracy and clarity. Further inspection indicates that this degradation is mainly caused by inconsistent compliance with output-format constraints, rather than deficiencies in content understanding. Apart from this structural issue, all models demonstrate relatively strong performance across the remaining criteria, and the overall trends remain consistent with those observed in data-query evaluations.

Fig. \ref{figs4}(c) provides an aggregated view across all five evaluation dimensions for comprehensive queries. As model scale increases, performance improves uniformly across dimensions, further confirming the positive correlation between parameter size and overall capability. Notably, the mid-sized model that underperforms in structure quality for direct data and knowledge queries behaves more consistently in comprehensive-query settings, where intermediate reasoning outputs are routed to a synthesis agent. This design mitigates formatting instability and leads to more reliable final responses.

Comparisons between human and hybrid expert evaluations reveal closely aligned scoring trends across all tasks and model scales, suggesting that the hybrid evaluation protocol provides a reliable approximation of human judgment. In summary, while larger models consistently yield higher-quality outputs, mid-sized models may exhibit occasional instability in output structure, which should be carefully considered in practical deployment scenarios.

\subsection*{Note S3: Procedures for selection of standard operations}
The selection procedure for identifying standard operations is outlined in Algorithm \ref{algo1}. For each operation, the corresponding time stamps are first extracted and evaluated against the duration threshold $T_{\min}$. Only operations that satisfy this requirement proceed to the subsequent stability assessment. Owing to the inherent current ramp-up and ramp-down behavior in BESS applications, only the middle segment of each current profile is considered for this assessment. The optimal fitting constant $c_i^\star$ is obtained using the Least Squares (LS) method. The resulting constant $c_i^\star$ and the associated error metric $\mathrm{RMSE}_i$ are then compared with the predefined thresholds $c_{\text{th}}$ and $\varepsilon_{\mathrm{RMSE}}$, respectively. In the studied cases, $T_{\min} = 5400$ s, $c_{\text{th}}=110$ A, and $\varepsilon_{\mathrm{RMSE}}=15$ are determined according to the operational characteristics of the studied BESS.

\setcounter{algocf}{0}
\renewcommand{\thealgocf}{\arabic{algocf}}
\begin{algorithm}[h]
\caption{Selection of standard operation cycles.}
\label{algo1}
\Initialize{Given operation set $\{\mathrm{Op}_i\}_{i=1}^N$, duration threshold $T_{\min}$, RMSE threshold $\varepsilon_{\mathrm{RMSE}}$, constant threshold $c_{\text{th}}$, standard operation set $\mathcal{S}_{\mathrm{std}} = \emptyset$}
\For{$i = 1, 2, \dots, N$}
{
  Extract current samples $\{I_i(t_j)\}_{j=1}^{n_i}$ and time stamps $\{t_j\}_{j=1}^{n_i}$ from $\mathrm{Op}_i$\;
  Compute total duration $T_i = t_{n_i} - t_1$\;
  \If{$T_i < T_{\min}$}
  {
    \textbf{continue}
  }
  Compute trimming length $\Delta T = (t_{n_i} - t_1)/6$\;
  Set $t_{\text{start}} = t_1 + \Delta T$, $t_{\text{end}} = t_{n_i} - \Delta T$\;
  Form middle index set $\mathcal{J}_i = \{ j \mid t_{\text{start}} \le t_j \le t_{\text{end}} \}$\;
  Compute
    $c_i^\star = \arg\min_{c \in \mathbb{R}}
    \frac{1}{|\mathcal{J}_i|}\sum_{j \in \mathcal{J}_i}\bigl(I_i(t_j)-c\bigr)^2$\;
  \If{$c_i^\star < c_{\text{th}}$}
  {
    \textbf{continue}
  }
  Compute $\mathrm{RMSE}_i =
    \sqrt{\dfrac{1}{|\mathcal{J}_i|}
    \sum_{j \in \mathcal{J}_i} \bigl(I_i(t_j) - c_i^\star\bigr)^2 }$\;
  \If{$\mathrm{RMSE}_i \le \varepsilon_{\mathrm{RMSE}}$}
  {
    $\mathcal{S}_{\mathrm{std}} \gets \mathcal{S}_{\mathrm{std}} \cup \{\mathrm{Op}_i\}$\;
  }
}
\Ret{$\mathcal{S}_{\mathrm{std}}$}
\end{algorithm}

\newpage

\subsection*{Note S4: Details for inconsistency evaluation methods}

After selecting the standard operating conditions, a comprehensive evaluation is conducted to obtain a holistic characterization of the battery pack's inconsistency with the following three perspectives. 

Electrical inconsistency can lead to hazardous states such as overcharging or over-discharging, thereby threatening both performance and safety \cite{voltage_incon}. Thermal inconsistency often appears as localized overheating, elevated cooling requirements, and diminished energy efficiency \cite{thermal_incon}. Aging inconsistency arises from uneven capacity fade among cells, which directly compromises the long-term reliability of the pack \cite{aging_incon}. Inspired by previous work \cite{diagnosing}, this paper integrates electrical, thermal, and aging factors to evaluate the overall inconsistency of the battery pack.

\textbf{Evaluation of voltage inconsistency: }From the voltage perspective, denote the measured cell-voltage data of one battery pack under a specific operating condition as \[
\mathbf{A} = [\mathbf{a}_1, \mathbf{a}_2, \ldots, \mathbf{a}_m] \in \mathbb{R}^{n\times m},
\] 
containing the time-series voltage measurements of cell $j$, 
$m$ is the number of cells in the pack, and $n$ is the number of samples. 

At the $i$-th sampling instant, the instantaneous voltage spread of all cells in the pack is defined as
\begin{equation}
r_i^{(V)} = \max_{1 \leq j \leq m} A_{ij} \;-\; \min_{1 \leq j \leq m} A_{ij}.
\end{equation}
Based on the sequence $\{ r_i^{(V)} \}_{i=1}^n$, two range-based features, including the maximum voltage range and the average voltage range, are extracted and represented as
\begin{equation}
\Delta V_{\mathrm{max}} = \max\limits_{1 \le i \le n} r_i^{(V)},
\end{equation}
\begin{equation}
\Delta V_{\mathrm{mean}}= \frac{1}{n} \sum_{i=1}^{n} r_i^{(V)}.
\end{equation}
The maximum voltage range $\Delta V_{\mathrm{max}}$ captures the most severe electrical imbalance observed during an operation, as extreme deviations are closely linked to safety-critical risks such as overcharge and over-discharge. The average voltage range $\Delta V_{\mathrm{mean}}$ reflects the mean imbalance level across the entire sampling period, representing the general inconsistency of the pack \cite{voltage_incon1}.

In addition to these two statistics, the number of cells classified as inconsistent is also included as an evaluation metric by using the low-rank subspace projection method, adapted from the previous work \cite{diagnosing}. The core idea is to construct a mapping from the cell's voltage matrix $\mathbf{A}$ to a category set representing normal or inconsistent state. Consequently, the number of inconsistent cells in the pack is denoted as $\mathrm{num}\in\mathbb{N}$.

The identification begins by extracting the latent voltage patterns. The voltage data matrix is first decomposed into a low-rank component and a sparse component through Robust Principal Component Analysis (RPCA) \cite{rpca}. The objective of RPCA is to identify a low-rank matrix $\mathbf{L}$ and a sparse matrix $\mathbf{S}$ that satisfy
\begin{equation}
\min_{\mathbf{L},\mathbf{S}} \, \text{rank}(\mathbf{L}) + \|\mathbf{S}\|_0 
\quad \text{s.t.} \quad 
\mathbf{L} + \mathbf{S} = \mathbf{A},
\label{opt1}
\end{equation}
where $\text{rank}(\cdot)$ denotes matrix rank and $\|\cdot\|_0$ denotes the $l_0$ norm. By invoking a convex relaxation of the above formulation and further adopting the augmented Lagrangian framework, the resulting optimization problem can be effectively solved using the Alternating Directions Method (ADM).

Following the decomposition, Singular Value Decomposition (SVD) is applied to the low-rank matrix $\mathbf{L}$, yielding $\mathbf{L} = \mathbf{U}\Sigma\mathbf{V}^\top$, where $\mathbf{V}$ denotes the right singular matrix. The first column of $\mathbf{V}$, denoted by $\mathbf{v}$, is extracted as the representative feature vector of $\mathbf{L}$. The original charging voltage matrix is subsequently projected onto $\mathbf{v}$, resulting in $\mathbf{f} = \mathbf{A}\mathbf{v}$, where $\mathbf{f}$ characterizes the inconsistency feature embedded in the charging data.

Finally, the vector $\mathbf{f}$ is normalized to obtain the inconsistency indicator for each cell:
\begin{equation}
f_i' = \frac{f_i - \mu_{\mathbf{f}}}{\sigma_{\mathbf{f}}},
\label{norm}
\end{equation}
where $f_i$ is the $i$th component of $\mathbf{f}$, $\mu_{\mathbf{f}}$ and $\sigma_{\mathbf{f}}$ denote the mean and standard deviation of $\mathbf{f}$, respectively. The normalized feature $f_i'$ encapsulates the inconsistency level of the $i$th cell, and the classification is performed according to Table below, where the criterion threshold is determined by both experience and reference from related works \cite{diagnosing, inconsistency_eva}.

\begin{table}[h]
\centering
\label{table2}
\begin{tabular}{ccc}
\toprule 
Inconsistency level  &  Criteria \\
\toprule 
Normal  & $f_i^\prime\leq4.5$\\
Inconsistent  & $f_i^\prime>4.5$\\
\bottomrule 
\end{tabular}
\end{table}

The identification result is illustrated in Fig. \ref{figs2} for reference.

\vspace{1em}

\textbf{Evaluation of thermal inconsistency:} From the thermal perspective, denote the measured temperature data of a battery pack under a specific operating condition as 
\[\mathbf{B} = [\, \mathbf{b}_1, \mathbf{b}_2, \ldots, \mathbf{b}_p \,] \in \mathbb{R}^{q \times p}  \],
where each column $\mathbf{b}_j \in \mathbb{R}^{q}$ is the time-series temperature measurements of sensor $j$, 
$p$ is the number of sensors in the pack, and $q$ is the number of samples. 

Similarly, at the $i$-th sampling instant, the instantaneous temperature difference of all sensors in the pack is defined as
\begin{equation}
r_i^{(T)} = \max_{1 \leq j \leq p} B_{ij} \;-\; \min_{1 \leq j \leq p} B_{ij}.
\end{equation}
Based on the sequence $r_i^{(T)}$, two range-based features, including maximum temperature difference and average temperature difference, are extracted and represented as
\begin{equation}
\Delta T_{\mathrm{max}} = \max\limits_{1 \le i \le q} r_i^{(T)},
\end{equation}
\begin{equation}
\Delta T_{\mathrm{mean}} = \frac{1}{q} \sum_{i=1}^{q} r_i^{(T)}.
\end{equation}
The maximum temperature difference $\Delta T_{\mathrm{max}}$ captures the most severe thermal imbalance observed during operation, which is closely related to critical safety risks such as local overheating and potential thermal runaway. The average temperature difference $\Delta T_{\mathrm{mean}}$ reflects the mean imbalance level across the entire sampling period, representing the general thermal inconsistency of the pack \cite{thermal_incon1}.

To further represent the thermal inconsistency within the battery pack, the Thermal Consistency Coefficient (TCC) is introduced. Based on the prior work, TCC can serve as an effective indicator of thermal inconsistency, reflecting local material or cooling variations and flagging issues like thermal degradation or cooling-system malfunctions \cite{diagnosing}. Compared with using only the maximum or average temperature range, TCC provides a time-integrated, sensor-aware, and normalization-based measure that more reliably captures persistent and structural thermal inconsistencies, rather than transient or sensor-specific fluctuations. It is formulated as:
\begin{equation}
\label{eq16}
\mathrm{TCC} = \sum_{t=2}^{q} \frac{\sum_{j=1}^p \left( B_{tj} - B_{1j} \right)}{p \times (t-1) \times \left( \max_j B_{tj} - \min_j B_{tj} \right)},
\end{equation}
where $B_{tj}$ is the temperature from sensor $j$ at sampling indices $t$.

\vspace{1em}

\textbf{Health evaluation: }From the health perspective, the health status of a battery pack reflects its degradation trajectory and directly affects BESS performance. Since in-service battery packs cannot be disassembled for inspection, explicit health labels are unavailable. Therefore, the Ampere-hour counting method is employed to calculate the capacity:
\begin{equation}
- \Delta t_i \sum_{k=k_1}^{k_2} \eta i[k] = Q\big(z[k_2] - z[k_1]\big),
\label{eq20}
\end{equation}
where $\Delta t_i$ is the sampling interval, $k_1$ and $k_2$ are discrete sampling indices corresponding to two different time instants, $\eta$ is the Coulomb efficiency, $i[k]$ is the pack current at time index $k$, $z[k]$ is the SOC at time index $k$, and $Q$ is the pack capacity.

Due to the existence of noise and fluctuations in both current and SOC measurements, only stable current segments are used for capacity estimation. Following the approach in \cite{diagnosing}, the Improved Recursive Least Squares (ImRLS) method is employed. Within this method, the Local Outlier Factor (LOF) algorithm is first applied to the current time series to identify steady charging/discharging periods, ensuring the validity of applying Equation \eqref{eq20}.

For each identified period, the left-hand side of Equation \eqref{eq20} is denoted as $y_i$ and the SOC difference as $x_i$, with $\sigma_{y_i}$ and $\sigma_{x_i}$ representing the variances associated with current measurement and SOC estimation, respectively. These $(x_i, y_i)$ pairs are used to formulate a Recursive Approximate Weighted Total Least Squares (RAWTLS) problem, which serves as the parameter estimation step of the ImRLS method. The corresponding cost function is given by:
\begin{equation}
\ell = \sum_{i=1}^{n} 
\frac{(y_i - \hat{Q} x_i)^2}{(1 + \hat{Q}^2)^2} 
\left( \frac{\hat{Q}^2}{\sigma_{x_i}^2} + \frac{1}{\sigma_{y_i}^2} \right),
\label{eq26}
\end{equation}
where $\hat{Q}$ denotes the estimated pack capacity. Minimization of Equation \eqref{eq26} via the RAWTLS procedure yields the capacity estimate $\hat{Q}$ \cite{ls}.

Finally, the health status of the battery pack is calculated as
\[
\mathrm{SOH} = \frac{\hat{Q}}{Q_{nom}},
\]
where $Q_{nom}$ is the nominal pack capacity.

\newpage

\subsection*{Note S5: Details for construction of knowledge dataset}
The knowledge dataset is built through procedures consisting of literature selection, knowledge distillation, and dataset construction.
\begin{itemize}
\item \textbf{Literature selection:} Relevant literature is gathered from target domains, including BESS applications, battery inconsistency mechanisms, and battery fault mechanisms. Review articles are prioritized over research articles.
They summarize general principles, mechanisms, and established findings across a broad scope, which makes them more suitable for knowledge-oriented query support. In this study, ten carefully curated review articles are selected as foundational materials, listed as follows.
\begin{enumerate}
\item A review of battery failure: classification, mechanisms, analysis, and management \cite{knowledge_dataset1}
\item Battery hazards for large energy storage
systems \cite{knowledge_dataset2}
\item Propagation mechanisms and diagnosis of parameter inconsistency within Li-Ion battery packs \cite{knowledge_dataset3}
\item Understanding the voltage inconsistency features in lithium-ion battery module \cite{voltage_incon}
\item Grid-connected battery energy storage system: a review on application and integration \cite{knowledge_dataset5}
\item How to better share energy towards a carbon-neutral city? A review on application strategies of battery energy storage system in city \cite{knowledge_dataset6}
\item A comprehensive review on inconsistency and equalization technology of lithium-ion battery for electric vehicles \cite{knowledge_dataset7}
\item Lithium-Ion Battery Pack Robust State of Charge
Estimation, Cell Inconsistency, and
Balancing: Review \cite{knowledge_dataset8}
\item Overcoming the challenges of integrating variable renewable energy to the grid: A comprehensive review of electrochemical battery storage systems \cite{knowledge_dataset9}
\item A critical review on inconsistency mechanism, evaluation methods and improvement measures for lithium-ion battery energy storage systems \cite{knowledge_dataset10}
\end{enumerate}

\item \textbf{Knowledge piece generation:}
Instead of directly slicing documents into fixed-length text chunks \cite{embedding}, an LLM-driven distillation strategy is adopted. Fixed-length slicing introduces two key limitations: it preserves irrelevant citation markers and mathematical expressions, and it disrupts semantic coherence by fragmenting conceptually continuous content \cite{llm_rag}. To address these issues, carefully designed prompts guide the LLM to extract and summarize essential information from each article, producing semantically coherent knowledge slices with controlled length. The prompt used for this process is provided below (corresponds to Prompt 1 shown in Fig. 2(b3)).

\begin{tcolorbox}[
title=\textcolor{black}{\textbf{Prompt 1}},
colback=gray!10,      
colframe=gray!40,     
boxrule=0.5pt,        
arc=2mm,              
left=6pt,right=6pt,top=6pt,bottom=6pt  
]
\small
\noindent\textbf{Role:} You are a domain expert in lithium-ion battery systems.

\noindent\textbf{Task:} Distill and segment the academic literature below into concise, retrieval-friendly knowledge slices.

\noindent\textbf{Objective:} Extract only the core conclusions, mechanisms, and quantified findings related to battery energy storage systems, battery inconsistency mechanisms, or battery failure mechanisms.

\noindent\textbf{Output format:}
\begin{itemize}
    \item The output must be in Markdown.
    \item Each knowledge slice must begin with a highly condensed one-sentence title, formatted as a Markdown heading using a leading \texttt{\#}.
    \item The slice content must appear directly under its title to clearly separate slices.
    \item Each slice should describe a clear cause-effect relationship, mechanism, or conclusion.
    \item The content length of each slice should be around 80-120 words.
\end{itemize}

\noindent\textbf{Constraints:}
\begin{itemize}
    \item Base the summary strictly on the provided text; do not introduce external facts or assumptions.
    \item Include numerical values, operating conditions, and thresholds whenever they appear.
    \item Avoid citations, equations, background theory, and references.
    \item Merge overlapping insights and remove redundancy.
    \item Do not use symbolic shorthand (e.g., delta notation); write terms in normal words.
\end{itemize}
\end{tcolorbox}

\item \textbf{Dataset construction:}
For each distilled knowledge slice, an additional semantic key is generated by the LLM to capture the core meaning of the content in a compact form. All slices and keys undergo manual review to ensure accuracy and relevance. Only the validated semantic keys, rather than the full slice texts, are embedded and stored in the vector database. During RAG, user queries are matched directly against these semantic keys, which improve retrieval accuracy and efficiency.
\end{itemize}

\newpage

\subsection*{Note S6: Details for the construction of multi-agent system}

Initially, Agent 1 performs intention analysis on the user's query. This step is necessary because different question types require different downstream procedures for accurate answering. Queries from BESS engineers are broadly classified into three categories: data queries, knowledge queries, and comprehensive queries.
\begin{itemize}
    \item The data query only requests data-specific extraction or analysis.
    \item The knowledge query only asks for specific expert knowledge.
    \item The comprehensive query requires both data analysis and expert knowledge to provide explanations or actionable suggestions.
\end{itemize}

Using a carefully designed Prompt 2 (corresponds to Prompt 2 shown in Fig. 2(b3)), Agent 1 determines an appropriate routing path for each query. 

\begin{tcolorbox}[
title=\textcolor{black}{\textbf{Prompt 2}},
colback=gray!10,      
colframe=gray!40,     
boxrule=0.5pt,        
arc=2mm,              
left=6pt,right=6pt,top=6pt,bottom=6pt  
]
\small

\noindent\textbf{Role:} You are a semantic router and query splitter for an Energy Storage System (ESS) inconsistency analysis pipeline.

A user question may contain: (1) requests to analyze historical ESS inconsistency data from an internal database (Branch-Data), (2) requests for general knowledge about mechanisms, causes, impacts, or optimization strategies (Branch-Knowledge), (3) or both.

\noindent\textbf{Task A: Routing path Determination}
\begin{itemize}
    \item Choose exactly one routing path by interpreting user intent:
    \begin{itemize}
        \item data\_only: The user requests numerical analysis of a specific dataset (time range, station, inconsistency indicators) without asking for mechanisms or optimization.
        \item knowledge\_only: The user requests conceptual ESS knowledge such as mechanisms, causes, impacts, standards, design considerations, or optimization strategies, without referencing any dataset.
        \item data\_and\_knowledge: The user references a dataset and also expects mechanism-level explanations or optimization suggestions.
    \end{itemize}
    \item Routing rules:
    \begin{itemize}
        \item If a time range is given and the user asks "why" or requests optimization, classify as data\_and\_knowledge.
        \item If a time range is given and the user only wants statistics, trends, or anomaly detection,  classify as data\_only.
        \item If no time range is given and the user asks only conceptual questions, classify as knowledge\_only.
        \item When ambiguous, prefer data\_and\_knowledge.
    \end{itemize}
\end{itemize}
\noindent\textbf{Task B: Generate Sub-Queries}
\begin{itemize}
    \item Produce two outputs: data\_query and knowledge\_query.
\end{itemize}
\end{tcolorbox}

\begin{tcolorbox}[
title=\textcolor{black}{\textbf{Prompt 2 (continue)}},
colback=gray!10,      
colframe=gray!40,     
boxrule=0.5pt,        
arc=2mm,              
left=6pt,right=6pt,top=6pt,bottom=6pt  
]
\small
\begin{itemize}
    \item data\_query rules:
    \begin{itemize}
        \item Include specific time ranges, station names, and references to inconsistency indicators.
        \item Specify that internal inconsistency datasets must be retrieved and numerically analyzed.
        \item Do not reference real-time or external data sources.
        \item If no data intent exists, set data\_query = "".
    \end{itemize}
    \item knowledge\_query rules:
    \begin{itemize}
        \item Remove dates and data retrieval requirements.
        \item Convert user intent into general ESS inconsistency questions (mechanisms, root causes, impacts, mitigation, optimization).
        \item Must focus on stationary ESS; EV-only topics should be excluded unless shared principles.
        \item If no knowledge intent exists, set knowledge\_query = "".
    \end{itemize}
\end{itemize}
\noindent\textbf{Required Output Format}
\begin{itemize}
    \item The final output must be valid JSON with no additional text:
\end{itemize}

\begin{lstlisting}[language=json, basicstyle=\ttfamily\footnotesize]
{
  "route": "data_only" | "knowledge_only" | "data_and_knowledge",
  "data_query": "...",
  "knowledge_query": "..."
}
\end{lstlisting}

\noindent\textbf{Examples}
\begin{lstlisting}[language=json, basicstyle=\ttfamily\footnotesize]
{
  "route": "data_only",
  "data_query": "Retrieve and analyze the internal inconsistency dataset of station A from 2025-04-01 to 2025-05-01 and compute voltage inconsistency trends and abnormal days.",
  "knowledge_query": ""
}
\end{lstlisting}
\begin{lstlisting}[language=json, basicstyle=\ttfamily\footnotesize]
{
  "route": "knowledge_only",
  "data_query": "",
  "knowledge_query": "What are the mechanisms, root causes, impacts, and optimization strategies for voltage and temperature inconsistency in ESS?"
}
\end{lstlisting}
\begin{lstlisting}[language=json, basicstyle=\ttfamily\footnotesize]
{
  "route": "data_and_knowledge",
  "data_query": "Retrieve the internal inconsistency dataset from 2025-04-01 to 2025-05-01 and perform numerical analysis on voltage and temperature inconsistency.",
  "knowledge_query": "What are the typical mechanisms, root causes, and optimization strategies for voltage and temperature inconsistency in ESS?"
}
\end{lstlisting}
\end{tcolorbox}

\noindent\textbf{Procedures for data query}

Once the query is identified as a data query, Agent 1 further extracts the specific temporal constraints embedded in the user request, such as the start and end dates of the data to be accessed.
Based on the extracted time range, the corresponding records from the database are extracted. Since each record is indexed by date in its title, the retrieval process can be efficiently implemented through direct date-based matching.
After the relevant data entries are obtained, Agent 2 generates the final response solely based on the retrieved data. This process consists of the following steps:
\begin{itemize}
\item \textbf{Data synthesis:} Integrate and consolidate information strictly based on the retrieved data records, without introducing any external or inferred data.
\item \textbf{Data analysis:} Reasoning operations, such as summarization, comparison, or trend detection, are then applied to the synthesized data to produce the final answer.
\end{itemize}
The above mentioned steps are conducted by proper prompt engineering illustrated in Prompt 3 (corresponds to Prompt 3 shown in Fig. 2(b3)).

\begin{tcolorbox}[
title=\textcolor{black}{\textbf{Prompt 3}},
colback=gray!10,      
colframe=gray!40,     
boxrule=0.5pt,        
arc=2mm,              
left=6pt,right=6pt,top=6pt,bottom=6pt  
]
\small
\begin{itemize}
  \item Here are the retrieved records from the database: \texttt{f"\{context\}"}
  \item Now here is the user question: \texttt{f"\{question\}"}
  \item You must perform the analysis based strictly on the available records.
\end{itemize}

\begin{itemize}
  \item Important rules you must follow:
  \begin{enumerate}
    \item You MUST analyze the available V, T, and H matrices even if the data covers only part of the user's requested date range.
    \item If the available records do not span the full date interval, you must still perform engineering analysis on the retrieved days.
    \item Trend analysis is NOT required unless the user explicitly asks for trends.
    \item If there are only one or two days of data, perform STATIC analysis such as:
    \begin{itemize}
      \item pack-to-pack voltage inconsistency comparison
      \item severity ranking of packs
      \item identification of abnormal packs or outliers
      \item comparison between the available days (if $>1$ day)
      \item interpretation of: V\_row1 (worst-case spread); V\_row2 (average spread); V\_row3 (bad cells)
    \end{itemize}
    \item You must NEVER answer that the analysis cannot be performed due to insufficient data unless the matrices themselves are missing.
    \item Provide a detailed, logically consistent engineering explanation using ONLY the retrieved records above.
    \item Do not mention missing days unless it directly affects interpretation.
  \end{enumerate}
\end{itemize}
\end{tcolorbox}

\begin{tcolorbox}[
title=\textcolor{black}{\textbf{Prompt 3 (continue)}},
colback=gray!10,      
colframe=gray!40,     
boxrule=0.5pt,        
arc=2mm,              
left=6pt,right=6pt,top=6pt,bottom=6pt  
]
\small
\begin{itemize}
  \item Now you MUST produce your result in the following STRICT JSON format:
 \begin{itemize}
  \item "data\_analysis": "Long, step-by-step engineering reasoning and numerical interpretation.",
  \item "data\_summary": "Detailed conclusion written in markdown, 2-6 short paragraphs, focusing on causes, severity, and optimization suggestions.",
  \item "data\_brief": "Very concise conclusion for on-site O\&M staff. Use ONLY 3-5 bullet points, each bullet is a single short sentence (<25 words), directly answering the user's question and giving actionable suggestions. The content MUST be valid markdown, starting each bullet with '- '. Do not include any other text outside the bullet list."
\end{itemize}
  \item Important format rules:
  \begin{itemize}
    \item You MUST output a single JSON object.
    \item Do NOT wrap the JSON in \texttt{```} or any code block.
    \item Do NOT add any extra comments or explanations outside the JSON.
  \end{itemize}
\end{itemize}
\noindent\textbf{Domain Context and Data Definitions}
\begin{itemize}
    \item You are a senior battery engineer and data analyst working with time-series inconsistency data that consists of three components: V (voltage inconsistency matrix), T (temperature inconsistency matrix), and H (SOH/health vector). These contain statistical features extracted from cell-level measurements for each battery pack.
    \item \textbf{Voltage inconsistency matrix (V)}:  
    Each column represents a battery pack.  
    Rows represent statistical features:  
    Row 1 is the maximum value of (max cell voltage minus min cell voltage), describing worst-case voltage spread; larger values mean worse internal consistency.  
    Row 2 is the mean value of this voltage difference, reflecting average spread; larger values also indicate worse consistency.  
    Row 3 is the count of severely inconsistent cells exceeding the internal inconsistency threshold; smaller values are preferred.
    \item \textbf{Temperature inconsistency matrix (T)}:  
    Each column represents a battery pack.  
    Rows represent thermal features:  
    Row 1 is the maximum difference between maximum and minimum cell temperatures, describing worst-case thermal non-uniformity; larger values indicate worse consistency.  
    Row 2 is the mean temperature difference; larger values represent worse thermal distribution.  
    Row 3 is a thermal consistency coefficient from 0 to 1, where higher values indicate better uniformity.
    \item \textbf{Health vector (H)}:  
    A single-row vector with each column representing a battery pack. SOH values range from 0 to 1; higher values represent healthier packs.
\end{itemize}
\noindent\textbf{Modeling and Reasoning Rules}
\begin{itemize}
    \item All columns always correspond to independent battery packs, and each row represents a distinct statistical feature extracted from time-series data.
    \item Reasoning must remain numerical, physically valid, and strictly grounded in the given definitions.
\end{itemize}
\end{tcolorbox}

\begin{tcolorbox}[
title=\textcolor{black}{\textbf{Prompt 3 (continue)}},
colback=gray!10,      
colframe=gray!40,     
boxrule=0.5pt,        
arc=2mm,              
left=6pt,right=6pt,top=6pt,bottom=6pt  
]
\small
\begin{itemize}
    \item When answering:
    \begin{itemize}
        \item Use only the retrieved Markdown slices as evidence.
        \item Show intermediate steps for all calculations.
        \item Explicitly state missing information when applicable.
        \item Do not invent data under any circumstances.
    \end{itemize}
    \item When the available data covers only part of the requested date range, full engineering analysis must still be carried out.
    \item Insufficient-data responses are allowed only if the matrices themselves are missing.
    \item Even minimal data (one or two days) must still be analyzed, including pack-to-pack differences, severity ranking, voltage spread distribution, anomalies, and comparisons across all available records.
\end{itemize}
\end{tcolorbox}

\vspace{2em}

\noindent\textbf{Procedures for knowledge query}

RAG is adopted as the core paradigm for knowledge query routing to ensure answer accuracy and factual grounding \cite{rag}. By querying an external knowledge dataset and injecting retrieved evidence into the language model, the system remains anchored in verified information while retaining the model's generalization ability. The overall workflow follows four sequential stages: query expansion, knowledge retrieval, expert reasoning, and relevance-driven knowledge integration. 

\begin{itemize}
\item \textbf{Query expansion:}
The original question is expanded into a compact set of sub-queries that cover key dimensions of BESS mechanism analysis, including root causes, system impacts, and mitigation strategies. This step improves retrieval robustness, particularly when the initial query is brief or underspecified \cite{querypara}.

\item \textbf{Knowledge retrieval:}
Each knowledge piece is organized as a slice consisting of a title key and an associated body text. The title key serves as a semantic summary for matching, while the body provides detailed evidence. Both expanded queries and keys are embedded into a normalized vector space using the FlagEmbedding model bge-base-en-v1.5 \cite{bge_m3}. Cosine similarity between the title key and expanded query is used to measure relevance. For each sub-query $q_i$, candidate keys $d_j$ are retrieved with similarity scores. Results across sub-queries are aggregated by retaining the maximum score per entry, and the Top-$k$ entries are selected. The detailed procedures are illustrated in Algorithm \ref{algo2}.

\begin{algorithm}[h]
\caption{Multi-query fusion retrieval via maximum cosine similarity.}
\label{algo2}
\Initialize{User question $q_0$; Set of query expansion $\{q_i\}_{i=1}^{n_1}$; Set of knowledge key $\{d_j\}_{j=1}^{n_2}$; Encoder for vectorization $\text{encoder}(\cdot)$; Number of retrieved entry $k$; Retrieved set $\mathcal{D}_{\text{top-}k} = \emptyset$}
\For{$i = 1, 2, \dots, n_1$}{
  Compute query embedding $\mathbf{q}_i = \text{encoder}(q_i)$ and normalize
  $\mathbf{q}_i \gets \mathbf{q}_i / \|\mathbf{q}_i\|_2$\;
}
\For{$j = 1, 2, \dots, n_2$}{
  Compute document embedding $\mathbf{d}_j = \text{encoder}(d_j)$ and normalize
  $\mathbf{d}_j \gets \mathbf{d}_j / \|\mathbf{d}_j\|_2$\;
}
\For{$j = 1, 2, \dots, n_2$}{
  Initialize fused score $s_j \gets -\infty$\;
  \For{$i = 1, 2, \dots, n_1$}{
    Compute cosine similarity
    $\mathrm{sim}_{ij} \gets \mathbf{q}_i^\top \mathbf{d}_j$\;
    \If{$\mathrm{sim}_{ij} > s_j$}{
      $s_j \gets \mathrm{sim}_{ij}$
    }
  }
}
Form index set
$\mathcal{J} \gets \{1,2,\dots,n_2\}$ and sort $\mathcal{J}$
such that $s_{j_1} \ge s_{j_2} \ge \dots \ge s_{j_{n_2}}$\;
Select top-$k$ indices
$\mathcal{J}_{\text{top-}k} \gets \{j_1, j_2, \dots, j_k\}$\;
\For{each $j \in \mathcal{J}_{\text{top-}k}$}{
  $\mathcal{D}_{\text{top-}k} \gets
    \mathcal{D}_{\text{top-}k} \cup \{d_j\}$\;
}
\Ret{$\mathcal{D}_{\text{top-}k}$}
\end{algorithm}

\item \textbf{Expert reasoning:}
Parallel to retrieval, Agent 3 produces a reasoning-based response derived solely from the language model's internal knowledge. This response captures engineering heuristics, domain expertise, and mechanistic understanding without relying on the external knowledge base. The detailed implementation can be referred from Prompt 4 (corresponds to Prompt 4 shown in Fig. 2(b3)) for expert reasoning.

\begin{tcolorbox}[
title=\textcolor{black}{\textbf{Prompt 4 for expert reasoning}},
colback=gray!10,
colframe=gray!40,
boxrule=0.5pt,
arc=2mm,
left=6pt,right=6pt,top=6pt,bottom=6pt
]
\small
\noindent\textbf{Role:} You are a senior Energy Storage System (ESS) engineer. You deeply understand electrochemical energy storage, battery pack design, thermal management, cooling systems, temperature consistency control, safety, and diagnostics. Your job is to directly answer the user's question based on your own knowledge, even if no external documents are available.

\noindent\textbf{Requirements:} 
\begin{itemize}
\item Focus strictly on ESS (stationary energy storage), unless the concept is clearly shared with EVs.
\item Give a clear, step-by-step technical explanation.
\item When appropriate, structure your answer into: background, root causes, analysis, and concrete recommendations.
\item If the question is too vague or missing key parameters, explicitly state the assumptions you are making.
\item If something is uncertain or design-dependent, say so explicitly instead of hallucinating specifics.
\end{itemize}
\end{tcolorbox}

\begin{tcolorbox}[
title=\textcolor{black}{\textbf{Prompt 4 for knowledge integration}},
colback=gray!10,
colframe=gray!40,
boxrule=0.5pt,
arc=2mm,
left=6pt,right=6pt,top=6pt,bottom=6pt
]
\small

\textbf{Role:} You are an expert reviewer and solution integrator for Energy Storage Systems (ESS).\\
You will be given:
\begin{itemize}
\item The user's question.
\item Retrieved knowledge snippets (RAG results).
\item A standalone expert answer.
\end{itemize}
Your tasks:
\begin{itemize}
\item 1) Judge the relevance of the RAG snippets as exactly one of: \texttt{"high"}, \texttt{"medium"}, or \texttt{"low"}.
\item 2) Produce a highly condensed, well-structured bullet summary for ESS O\&M staff, with constraints:
\begin{itemize}
\item Summary must contain ONLY 3-6 bullets.
\item Each bullet must be <25 words, start with \texttt{"- "}, valid markdown.
\item Bullets must follow:
\begin{itemize}
\item 1-2 bullets starting with [Mechanism]
\item 1-2 bullets starting with [Cause]
\item 1-2 bullets starting with [Mitigation]
\end{itemize}
\item Each bullet must end with exactly one tag: \texttt{[RAG]}, \texttt{[LLM]}, or \texttt{[RAG][LLM]}.
\item Bullets must be compact, engineering-focused, non-narrative.
\end{itemize}
\item 3) Output two compact views:
\begin{itemize}
\item \texttt{rag\_view}: conclusions supported mainly by RAG.
\item \texttt{llm\_view}: additions from standalone expert reasoning.
\end{itemize}
\item 4) Output exactly the following JSON object:
\begin{lstlisting}[language=json, basicstyle=\ttfamily\footnotesize]
{
"relevance": "high" | "medium" | "low",
"rag_view": "...",
"llm_view": "...",
"summary": "A markdown bullet list (3-6 bullets) following the rules above."
}
\end{lstlisting}
\end{itemize}
Rules:
\begin{itemize}
\item Do NOT output any text outside the JSON object.
\item Do NOT use code blocks.
\end{itemize}
\end{tcolorbox}

\item \textbf{Relevance-driven knowledge integration:}
Agent 3 evaluates the alignment between retrieved entries and the user query, assigning a relevance level of high, medium, or low. This level governs the integration policy. When relevance is high, the explanation is evidence-centric; when medium, retrieved evidence and expert reasoning are merged; when low, the system relies primarily on expert reasoning. The detailed implementation can be referred from Prompt 4 for knowlege integration.

\end{itemize}

To maintain scientific transparency, the final output annotates each statement with a source tag: sentences grounded mainly in retrieved evidence carry \texttt{[RAG]}, those based primarily on model reasoning carry \texttt{[LLM]}, and those synthesizing both appear as \texttt{[RAG][LLM]}. These tags provide a provenance mechanism indicating whether a statement stems from empirical knowledge entries, internal model inference, or a structured combination.

\vspace{2em}

\noindent\textbf{Procedures for comprehensive query}

For comprehensive query routing, the process integrates the structure of both data query routing and knowledge query routing, functioning as a synthesizer that fuses all available information. 
Once a query is classified as comprehensive, the system coordinates multiple agents to decompose, process, and fuse heterogeneous results into a coherent response.

\begin{itemize}

\item \textbf{Query decomposition.}
After the intention decision, Agent 1 splits the user's original query into two focused sub-queries: one targeting structured data analysis and the other targeting knowledge-based reasoning. This decomposition is implemented using Prompt 2 mentioned above.

\item \textbf{Data query processing.}
Following the predefined data routing pipeline, Agent 2 performs data synthesis and analysis, producing a concise data analysis summary that captures key quantitative patterns and results.

\item \textbf{Knowledge query processing.}
Following the predefined knowledge routing pipeline, Agent 3 generates a synthesized explanation grounded in retrieved evidence and expert reasoning.

\item \textbf{Cross-source synthesis.}
Acting as the synthesizer, Agent 4 integrates the original user query, the data analysis summary from Agent 2, and the knowledge synthesis output from Agent 3. It then generates a cohesive and well-structured final answer that jointly reflects data-driven insights and knowledge-based reasoning. The corresponding prompt can be referred to from Prompt 5 (corresponds to Prompt 5 shown in Fig. 2(b3)).
\end{itemize}

\begin{tcolorbox}[
title=\textcolor{black}{\textbf{Prompt 5}},
colback=gray!10,      
colframe=gray!40,     
boxrule=0.5pt,        
arc=2mm,              
left=6pt,right=6pt,top=6pt,bottom=6pt  
]
\small
\begin{itemize}
    \item The prompt provides retrieved database records and the user question. All analysis must be based strictly on these records. 
    \item You must analyze the available V, T, and H matrices even when the retrieved data covers only part of the user's requested date range. Full interval coverage is not required. 
    \item When the retrieved data does not span the full time range, engineering analysis must still be performed on the available days without rejecting the query. 
    \item Trend analysis should only be performed if the user explicitly asks for trends. Otherwise, conduct static evaluation. 
    \item If only one or two days of data are available, perform pack-level static assessments such as voltage inconsistency comparison, severity ranking, abnormal pack identification, multi-day comparison when applicable, and interpretation of V\_row1, V\_row2, and V\_row3. 
    \item You must not claim insufficient data unless the matrices themselves are missing. The answer must always include detailed, logically consistent engineering reasoning using only the retrieved records. 
    \item Missing days should not be mentioned unless they directly affect interpretation. 
    \item The final output must follow a strict JSON structure containing concise on-site O\&M guidance with exactly 3-5 bullets, each a single short actionable sentence under 25 words, valid Markdown beginning with "-".
    \item The output must be a single JSON object, without code fences or any extra commentary.
\end{itemize}

\end{tcolorbox}

\newpage

\section*{Supplementary Table}

\begin{table}[h]
\centering
\caption{Tested questions for data query}
\footnotesize
\begin{tabular}{|c|m{15cm}|}
\hline
\textbf{Index} & \textbf{Query} \\
\hline
1 & From 2024-10-01 to 2024-12-30, please list the days with the highest voltage spread of all packs. \\
\hline
2 & Retrieve the records between 2024-12-01 to 2025-01-30 and evaluate the thermal inconsistency of pack 7. \\
\hline
3 & For the period 2025-01-05 to 2025-04-20, summarize the health status of all packs. \\
\hline
4 & Query the dataset from 2025-04-01 to 2025-04-30 and identify all packs whose voltage imbalance exceeded 0.12 V. \\
\hline
5 & Between 2024-10-01 and 2025-01-01, analyze the average voltage spread of pack 1. \\
\hline
6 & Retrieve all records from 2025-01-01 to 2025-03-01 and find packs with the best thermal consistency. \\
\hline
7 & From 2025-01-10 to 2025-03-28, evaluate the health situation of pack 1. \\
\hline
8 & Please extract data from 2024-11-01 to 2024-11-30 and compare voltage inconsistency among all packs. \\
\hline
9 & Using records from 2025-02-01 to 2025-05-07, identify packs whose temperature spread exceeds 5 degrees. \\
\hline
10 & Retrieve and summarize all thermal inconsistency recorded between 2025-02-01 and 2025-05-01. \\
\hline
\end{tabular}
\label{table3}
\end{table}

\begin{table}[h]
\centering
\caption{Tested questions for knowledge query}
\footnotesize
\begin{tabular}{|c|m{15cm}|}
\hline
\textbf{Index} & \textbf{Query} \\
\hline
11 & What are the main mechanisms causing voltage inconsistency in large-scale ESS battery packs? \\
\hline
12 & How does temperature non-uniformity influence SOH degradation in energy storage systems? \\
\hline
13 & What design strategies can be used to improve thermal uniformity in liquid-cooled ESS modules? \\
\hline
14 & What are the typical root causes of severe cell-to-cell voltage spread in LFP-based ESS? \\
\hline
15 & How does poor thermal consistency affect safety and thermal runaway risk in stationary ESS? \\
\hline
16 & What operational practices help reduce voltage imbalance in long-term ESS cycling? \\
\hline
17 & Explain how voltage imbalance develops and propagates in multi-pack ESS clusters. \\
\hline
18 & How to prevent thermal inconsistency in BESS? \\
\hline
19 & How to improve the voltage consistency in battery packs when cell-to-cell deviation happens? \\
\hline
20 & How can SOH inconsistency across packs be mitigated in utility-scale energy storage systems? \\
\hline
\end{tabular}
\label{table4}
\end{table}

\newpage

\begin{table}[h]
\centering
\caption{Tested questions for comprehensive query}
\footnotesize
\begin{tabular}{|c|m{15cm}|}
\hline
\textbf{Index} & \textbf{Query} \\
\hline
21 & From 2025-01-01 to 2025-03-01, analyze the voltage inconsistency of all packs and explain the likely causes based on ESS inconsistency mechanisms. Provide optimization suggestions. \\
\hline
22 & Using the records from 2024-10-01 to 2024-12-01, evaluate thermal inconsistency and determine whether the patterns indicate cell aging or cooling system imbalance. \\
\hline
23 & Analyze the temperature and voltage inconsistency between 2025-02-01 and 2025-03-20 and explain what engineering factors may have contributed to the observed behavior. \\
\hline
24 & From 2025-01-07 to 2025-03-01, review the data and provide a root-cause explanation together with actionable recommendations to improve pack voltage uniformity. \\
\hline
25 & Between 2024-10-10 and 2024-11-25, identify the most voltage inconsistent packs and explain the possible mechanisms responsible. \\
\hline
26 & Using the dataset from 2025-03-01 to 2025-05-15, analyze the health status of pack 4 and describe what operational adjustments could ease aging. \\
\hline
27 & From 2024-12-01 to 2025-01-01, assess voltage and temperature inconsistency and provide engineering interpretation on the potential root causes. \\
\hline
28 & Analyze all records from 2025-02-01 to 2025-02-28 and propose both short-term and long-term optimization strategies. \\
\hline
29 & Using data from 2024-11-01 to 2024-12-01, evaluate pack 6 inconsistency and discuss what underlying mechanisms may be driving the observed variation. \\
\hline
30 & From 2025-04-01 to 2025-05-01, analyze the voltage, thermal and health status and explain what might be causing the inconsistencies, along with improvement recommendations. \\
\hline
\end{tabular}
\label{table5}
\end{table}

\bibliographystyle{elsarticle-num} 
\bibliography{supplement_citation}